\documentclass[12pt, draftclsnofoot, onecolumn]{IEEEtran}
\ifCLASSINFOpdf
\else
\usepackage[dvips]{graphicx}
\fi
\usepackage[cmex10]{amsmath}
\usepackage{upgreek}
\usepackage{booktabs}
\usepackage{epsfig}
\usepackage{latexsym}
\usepackage{multirow}
\usepackage{stfloats}
\usepackage{epstopdf}
\usepackage{color}  
\usepackage{tabularx} 
\usepackage{amssymb}
\usepackage{enumerate}
\graphicspath{{./Figures/}}
\usepackage{bbm}
\usepackage{bm}
\usepackage{cite}
\usepackage[tight,footnotesize]{subfigure}
\usepackage{balance}
\usepackage{mathrsfs}
\usepackage{verbatim}
\usepackage{dsfont}
\usepackage{verbatim}
\usepackage{tikz}
\usepackage{setspace}
\usepackage{diagbox}
\usepackage{caption}
\usepackage[framemethod=tikz]{mdframed}
\usepackage{multicol}
\usepackage{environ}
\usepackage{tikz}

\begin{document}
\title{
Dynamic-subarray with Fixed Phase Shifters for Energy-efficient Terahertz Hybrid Beamforming under Partial CSI
}
\author{
	Longfei Yan,
	Chong Han,~\IEEEmembership{Member,~IEEE}, Nan Yang,~\IEEEmembership{Senior Member,~IEEE}, \\
	and Jinhong Yuan,~\IEEEmembership{Fellow,~IEEE}
	\thanks{
	This paper was presented in part at the IEEE Globecom, December 7-11, 2020~\cite{DS_GC_2020}.
		
	Longfei Yan and Chong Han are with the Terahertz Wireless Communications (TWC) Laboratory, Shanghai Jiao Tong University, Shanghai 200240, China (e-mail: \{longfei.yan, chong.han\}@sjtu.edu.cn).
		
	Nan Yang is with the School of
	Engineering, Australian National University, Canberra, ACT 2600, Australia
	(e-mail: nan.yang@anu.edu.au).
		
	Jinhong Yuan is with the School of Electrical Engineering and Telecommunications, University of New South Wales, Sydney, NSW 2052, Australia (e-mail: j.yuan@unsw.edu.au).
}}
\maketitle
\markboth{}{}
\thispagestyle{empty} 
\begin{abstract}
\boldmath
Terahertz (THz) communications are regarded as a pillar technology for the 6G systems, by offering multi-ten-GHz bandwidth. To overcome the huge propagation loss while reducing the hardware complexity, THz ultra-massive (UM) MIMO systems with hybrid beamforming are proposed to offer high array gain. Notably, the adjustable-phase-shifters considered in most existing hybrid beamforming studies are power-hungry and difficult to realize in the THz band. Moreover, due to the ultra-massive antennas, full channel-state-information (CSI) is challenging to obtain. 
To address these practical concerns, in this paper, an energy-efficient dynamic-subarray with fixed-phase-shifters (DS-FPS) architecture is proposed for THz hybrid beamforming. To compensate for the spectral efficiency loss caused by the fixed-phase of FPS, a switch network is inserted to enable dynamic connections. In addition, by considering the partial CSI, we propose a row-successive-decomposition (RSD) algorithm to design the hybrid beamforming matrices for DS-FPS. A row-by-row (RBR) algorithm is further proposed to reduce computational complexity. 
Extensive simulation results show that, the proposed DS-FPS architecture with the RSD and RBR algorithms achieves much higher energy efficiency than the existing architectures. 
Moreover, the DS-FPS architecture with partial CSI achieves 97\% spectral efficiency of that with full CSI.
\end{abstract}
\IEEEpeerreviewmaketitle
\section{Introduction}
\label{section_intro}
In order to meet the rapid growth of wireless data rates, the Terahertz (THz) band with ultra-broad bandwidth has gained increasing attention~\cite{Nagatsuma2016Advances}.
In the THz band, the ultra-wide bandwidth comes with a cost of huge propagation loss, which drastically limits the communication distance~\cite{8387211}. Fortunately, the sub-millimeter wavelength allows the design of array consisting of a large number of antennas at transceivers, e.g., 1024, to enable THz ultra-massive MIMO (UM-MIMO) systems~\cite{AKYILDIZ201646}.
By utilizing the ultra-massive antennas, the beamforming technology can provide a high array gain to compensate for the path loss and combat the distance problem. Meanwhile, multiple data streams can be supported to offer a multiplexing gain and further improve the spectral efficiency of THz UM-MIMO systems~\cite{THz_HBF_WCM_2021}. 
In the THz UM-MIMO systems, many hardware constraints preclude from using conventional digital beamforming, which, instead, motivates the appealing hybrid beamforming~\cite{9398864,9411813}. The hybrid beamforming divides signal processing into the digital baseband domain and analog radio-frequency (RF) domain, which can achieve high spectral efficiency while maintaining low hardware complexity. 

The fully-connected (FC) and array-of-subarrays (AoSA) architectures are two widely-studied hybrid beamforming architectures~\cite{7436794,9139316,1,7397861,7913599,7445130}. The FC architecture achieves high spectral efficiency while consuming high power. On the contrary, as illustrated in Fig.~\ref{architecture_DS_FPS}(a), the complexity and power consumption of AoSA are noticeably reduced, while the spectral efficiency is largely sacrificed. As a result, the \textit{energy efficiency}, which is defined as the ratio between the spectral efficiency and power consumption, of both these two architectures is unsatisfactory and needs to be enhanced.
Moreover, in THz UM-MIMO systems, due to the large number of antennas and the high carrier frequency, full channel state information (CSI) is difficult to acquire.
To address these two practical concerns, we develop a novel energy-efficient hybrid beamforming architecture for THz UM-MIMO systems based on partial CSI in this work.
\subsection{Related Work}
\subsubsection{Energy-efficient hybrid beamforming architecture}
\begin{figure*}
	\centering
	\includegraphics[scale=0.32]{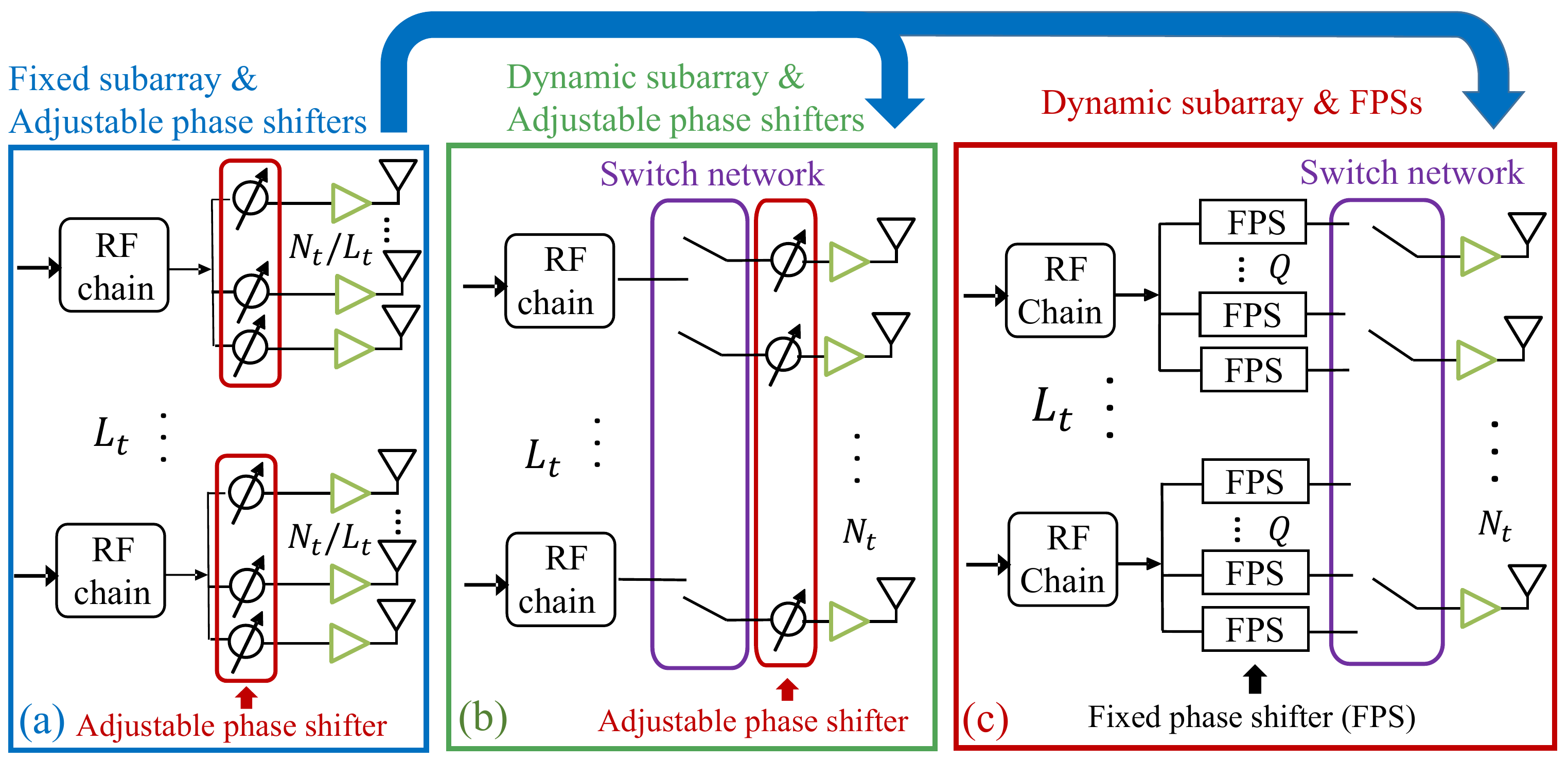} 
	\captionsetup{font={footnotesize}}
	\caption{The analog part of the architecture of hybrid beamforming. (a) AoSA architecture (b) DS architecture (c) The proposed DS-FPS architecture.
	}  
	\label{architecture_DS_FPS} 
	\vspace{-9.5mm}
\end{figure*}

There have been many studies aiming at improving the energy efficiency of the hybrid beamforming systems. The authors in~\cite{9205899} propose to jointly optimize the hybrid beamforming matrices and the resolution of DAC/ADC to enhance the energy efficiency. Another promising direction is utilizing the low-cost switches to provide dynamic connections. In~\cite{8778669,Aryan_paper1}, the RF chains of the hybrid beamforming architecture are dynamically deactivated to reduce the power consumption and enhance the energy efficiency.
Except from the RF chain selection, there have also been many efforts on the dynamic connections between phase shifters and antennas~\cite{DAoSA_JSAC_2020,9026753,7880698,8642953,9219133,9110865}.
By inserting switches in the FC architecture, a dynamic AoSA (DAoSA) architecture~\cite{DAoSA_JSAC_2020} and a fully-adaptive-connected (FAC) architecture~\cite{9026753} were proposed, where partial phase shifters are inactive to reduce power consumption. However, the number of remaining active phase shifters is usually larger than the number of antennas, which causes high power consumption. 
A dynamic hybrid beamforming (DHB) architecture was proposed~\cite{7880698,8642953}, as shown in Fig.~\ref{architecture_DS_FPS}(b). Through the switch network, each antenna dynamically selects one RF chain to connect with to enhance the spectral efficiency. Moreover, the number of phase shifters in the DHB architecture equals the number of antennas, which is less than the DAoSA and FAC architectures and leads to higher energy efficiency.

One remaining problem of the DHB architecture is that the phase shifters are assumed to own high resolution and even infinite resolution, which are impractical and power-hungry. To address this problem, the authors of~\cite{9219133,9110865} proposed to use low-resolution phase shifters in the DHB architecture.
However, both the above high-resolution and low-resolution phase shifters in~\cite{DAoSA_JSAC_2020,9026753,7880698,8642953,9219133,9110865} are still \textit{adjustable phase shifters}, i.e., phase selection is adjustable, which has high power consumption in the THz band. Fortunately, a \textit{fixed phase shifter} (FPS) can be adopted, whose phase remains fixed and non-adjustable. Compared to adjustable phase shifters, the THz FPS has substantially lower power consumption, which is more practical.
The authors of~\cite{7387790,8310586} proposed to use FPSs, instead of adjustable phase shifters, in the hybrid beamforming architecture, where each antenna is connected with multiple FPSs through switches. As a result, the number of closed switches is usually several times of the number of antennas, which is unbearably large in THz UM-MIMO systems and causes huge power consumption. Therefore, to utilize low-cost FPSs while keeping low power consumption, we propose a novel DS-FPS architecture, as shown in Fig.~\ref{architecture_DS_FPS}(c). In the proposed DS-FPS architecture, each RF chain connects with multiple FPSs. Each antenna can dynamically select one FPS through one switch such that the number of closed switches equals the number of antennas, which is much smaller than the number of closed switches in~\cite{7387790,8310586}. As a result, the energy efficiency of the proposed DS-FPS architecture is improved.

\subsubsection{Partial CSI}
Most of the existing hybrid beamforming studies assume that full CSI is known at both transmitter and receiver. However, the full CSI is hard to obtain in THz UM-MIMO systems due to the large-dimensional channel matrix caused by ultra-massive antennas. To tackle this problem, the authors of~\cite{7737056} considered to design the hybrid beamforming based on partial elements of the channel matrix. Since the overall dimension of the channel matrix in THz UM-MIMO systems is prohibitively large, acquiring partial elements of the channel matrix is sill difficult.
The authors of~\cite{8642953,7572969,9107073} proposed the hybrid beamforming solutions considering the statistical information of the channel. However, due to the lack of well-known general statistical MIMO channel model in the THz band yet, it is also challenging to know the statistical channel information.
A more practical partial CSI was considered in~\cite{7908940,6522603}, where only the directions and the amplitude of the path gains of the multipath are known. 
However, the hybrid beamforming algorithms in~\cite{7908940,6522603} were proposed for the FC architecture with adjustable phase shifters and did not consider the use of low-cost FPSs, thus cannot be applied to the proposed DS-FPS architecture in this work. 
Consequently, by considering the practical partial CSI, novel hybrid beamforming algorithms need to be proposed for the DS-FPS architecture.
\subsection{Our Contributions} 
In this work, we propose an energy-efficient DS-FPS architecture, by utilizing the low-cost FPS and dynamic switch network, as shown Fig.~\ref{architecture_DS_FPS}(c). Moreover, we consider the practical partial CSI, i.e., only the directions and the amplitude of the path gains of the multipath are known. Since the number of multipath of THz channel is usually around $5$~\cite{6998944}, the number of required parameters in the considered partial CSI is very limited. Furthermore, by considering partial CSI, we propose two CSI-robust hybrid beamforming algorithms for the THz DS-FPS architecture.
In the prior and shorter version of this work~\cite{DS_GC_2020}, we concisely investigated the DS-FPS architecture, while the consideration of partial CSI, the corresponding hybrid beamforming algorithms, and the performance comparisons with existing work in terms of spectral and energy efficiencies were not thoroughly studied.
The distinctive features of this work are summarized as follows.	
\begin{itemize}
    \item
    \textbf{We propose an energy-efficient DS-FPS architecture, by using the low-cost FPS and switch network.} With a fixed and nonadjustable phase, the FPS is more practical and consumes less power than the adjustable phase shifter in the THz band, which however, brings spectral efficiency loss. 
    To address this problem, we further design a switch network to enable dynamic connections between the antennas and FPSs. Each antenna can intelligently select one FPS with the proper phase from all FPSs to adapt the THz UM-MIMO channel, which enhances the spectral efficiency. 
	\item
	\textbf{By considering partial CSI, i.e., the directions and amplitude of path gains of multipath propagation, we formulate the hybrid beamforming problem for the DS-FPS architecture and propose two CSI-robust hybrid beamforming algorithms.} Specifically, we first propose a row-successive-decomposition (RSD) algorithm. The key idea is deriving an approximated form of the spectral efficiency, which only relies on partial CSI, and then optimizing each row of the switch network matrix successively. Furthermore, to reduce the computational complexity brought by the successive design, we propose a row-by-row (RBR) algorithm, which decomposes the optimization of each row of the switch network matrix as multiple uncorrelated sub-problems and solves them in parallel.
	\item
	\textbf{We evaluate the performance of the proposed DS-FPS architecture with the RSD and RBR algorithms and analyze the computational complexity.} Specifically, we show that the DS-FPS architecture achieves significantly higher energy efficiency than the existing architectures. Moreover, with partial CSI, the spectral efficiency of the RSD and RBR algorithms is similar to the case of full CSI and is robust to the CSI error. Furthermore, we analyze the computational complexity of the RBR algorithm, which is much lower than the existing algorithms. Compared to the RBR algorithm, the RSD algorithm yields a higher spectral efficiency, at the cost of increased computational complexity.
\end{itemize}  
	
The remainder of this paper is organized as follows. In Sec.~\ref{section_channel_system_model}, we present the channel model and system model, and formulate the hybrid beamforming problem for the THz DS-FPS architecture. Then, an RSD algorithm and a low-complexity RBR algorithm are proposed to solve the DS-FPS hybrid beamforming problem in Sec.~\ref{section_RSD_algorithm} and Sec.~\ref{section_RBR_algorithm}, respectively. Furthermore, simulation results are provided in Sec.~\ref{section_simulation}. Finally, the conclusion is drawn in Sec.~\ref{section_conclusion}.
	
\textbf{Notations}: $\textbf{A}$ is a matrix, $\textbf{a}$ is a vector, $a$ is a scalar.
$\textbf{I}_{N}$ denotes an $N$-dimensional identity matrix. $(\cdot)^T$, $(\cdot)^*$, and $(\cdot)^{H}$ represent transpose, conjugate, and conjugate transpose. $\lVert\cdot\rVert_{p}$ is the $p$-norm of the vector. $\lVert\cdot\rVert_{F}$ is the Frobenius norm of the matrix. ${\rm Tr}(\cdot)$ and ${\rm Re}(\cdot)$ denote the trace and real part of the matrix. ${\rm blkdiag}(\cdot)$ denotes the block diagonal matrix. $\odot$ is the element-wise product. $\otimes$ represents the Kronecker product.

\section{Channel Model and System Model of THz DS-FPS Hybrid Beamforming}
\label{section_channel_system_model}
In this section, we first introduce the THz channel model and the consideration of partial CSI.
Then, we investigate the system model with the DS-FPS architecture. Based on the considered partial CSI, we formulate the hybrid beamforming design problem of the DS-FPS architecture.

\subsection{Channel Model and the Practical Partial CSI}
We consider a wideband multi-carrier THz UM-MIMO system, where $k=1$, $2$, ... $K$ denotes the index of the sub-carrier. $N_t$ and $N_r$ represent the numbers of antennas at the transmitter and receiver, respectively.
The THz channel is usually very sparse, i.e., with limited multipath components~\cite{6998944}. Hence, we adopt a multipath model which is usually used for sparse MIMO channel as follows.
For the $k^{\rm th}$ sub-carrier whose frequency and wavelength are $f_k$ and $\lambda_k$, the channel matrix $\textbf{H}[k]\in\mathbb{C}^{N_r\times N_t}$ can be written as~\cite{6998944,9398864}
\begin{subequations}
\begin{align}
\textbf{H}[k]&=\sum\nolimits_{i=1}^{N_p}\!\alpha_{i}[k]
\textbf{a}_{ri}[k]\textbf{a}_{ti}[k]^{H}\label{channel_model_planar_1}\\
&=\textbf{A}_{r}[k]\bm{\Lambda}[k]\textbf{A}_{t}[k]^H
\label{channel_model_planar_2}\\
&=\textbf{A}_{r}[k]\left(\bm{\bar\Lambda}[k]\odot e^{j*\bm{\dot\Lambda}[k]}\right)\textbf{A}_{t}[k]^H,
\label{channel_model_planar_3}
\end{align}
\label{channel_model_planar}%
\end{subequations}
where $\alpha_{i}[k]$ describes the complex path gain of the $i^{th}$ path, $\textbf{a}_{ri}[k]$ and $\textbf{a}_{ti}[k]$ denote the received and transmitted array response vectors for the $i^{\rm th}$ path of the $k^{\rm th}$ sub-carrier, which are the $i^{\rm th}$ column of $\textbf{A}_r[k]$ and $\textbf{A}_t[k]$, respectively.
Additionally, $\bm{\Lambda}[k]\in\mathbb{C}^{N_p\times N_p}$ is a diagonal matrix, whose element at the $i^{\rm th}$ row and $i^{\rm th}$ column is $\alpha_{i}[k]$. We use $\bm{\bar\Lambda}[k]$ and $\bm{\dot\Lambda}[k]$ to denote the amplitude and the phase of $\bm{\Lambda}[k]$, respectively.
In this work, we consider a THz system at 0.3 THz. Specifically, we use the ray-tracing method to generate the directions, interactions with objects, and propagation distances of each propagation path, as shown in Sec.~\ref{section_simulation}-A. Then, we compute the path gains according to our previous THz channel work~\cite{6998944}, by considering the THz-specific propagation characteristics, including i) the high spreading loss, ii) the strong molecular absorption phenomena that renders severe frequency selectivity and the resulting temporal broadening effects, iii) strong penetration and rough-surface scattering, among others. As a result, the considered THz channel is different with mmWave channels.

We describe the expression for $\textbf{a}_{ri}[k]$ in \eqref{steering_vector_UPA}, while the expression for $\textbf{a}_{ti}[k]$ is similar and extensible. For an $L\times W$-element uniform planar array on the yz-plane, $\textbf{a}_{ri}[k]$ can be written as
\begin{equation}
	\textbf{a}_{ri}[k]=\big[1,
	... ,e^{j\frac{2\pi}{\lambda_k}d(L-1){\rm sin}(\phi_{ri}){\rm sin}(\theta_{ri})}\big]^T\otimes\big[1,
	... ,e^{j\frac{2\pi}{\lambda_k}d(W-1){\rm cos}(\theta_{ri})}\big]^T,
	\label{steering_vector_UPA}%
\end{equation}
where $\phi_{ri}$ and $\theta_{ri}$ are the azimuth and elevation direction of arrival (DoA) of the $i^{\rm th}$ path. For $\textbf{a}_{ti}[k]$, the angles $\phi_{ti}$ and $\theta_{ti}$ denote the azimuth and elevation direction of departure (DoD). Moreover, $d$ is the antenna spacing, which is half of the wavelength of the central frequency. 

\subsubsection{Partial CSI}
The channel matrix $\textbf{H}[k]$ is composed by the DoA, DoD, and path gain of each path.
There have been many studies that jointly estimate the DoA, DoD, and the path gain at either transmitter or receiver~\cite{7400949}. After the feedback, both the transmitter and receiver know the DoA, DoD, and path gain. In this work, we consider the partial CSI scenario, where the transmitter knows the DoD and amplitude of path gain, i.e., $\textbf{A}_t[k]$ and $\bm{\bar\Lambda}[k]$ in~\eqref{channel_model_planar_3}, while the receiver knows the DoA and amplitude of path gain, i.e., $\textbf{A}_r[k]$ and $\bm{\bar\Lambda}[k]$, respectively. There have been multiple low-complexity methods~\cite{7914742,8949442} which can estimate the DoA at the receiver and estimate the DoD at the transmitter, respectively. During the estimation of DoA and DoD, the amplitude of path gain can also be acquired. Consequently, the considered partial CSI in this work is practical.
For sparse channels, e.g., the THz channel in this work, the DoD and DoA are demanding information for hybrid beamforming design. While for rich scattering channels, it has been studied in~\cite{8647690} that the DoD and DoA can be mapped to the correlation matrix which is a realistic requirement for the hybrid beamforming design to achieve high spectral efficiency.

\subsection{System Model of DS-FPS Hybrid Beamforming}
Most of the existing mmWave hybrid beamforming studies~\cite{1,7397861,7913599,7445130,9026753,7880698,8642953,9219133,9110865} have used adjustable phase shifters, whose quantity is usually proportional to the number of antennas. For communications in the THz band, due to the higher frequency and larger number of antennas, the power consumption of adjustable phase shifters becomes prohibitively high and thus, impractical to use. To address this, we use low-cost FPSs to construct a DS-FPS architecture, as shown in Fig.~\ref{architecture_DS_FPS}(c). As will be shown in Sec.~\ref{section_simulation}, the DS-FPS architecture has much lower power consumption and higher energy efficiency than the existing adjustable phase shifters-based architectures, which indeed tackles the power consumption challenge of the THz band, although its spectral efficiency is lower than some existing architectures, e.g., the FC architecture. However, at mmWave systems, since the power consumption of adjustable phase shifters is still acceptable, as considered by most existing studies~\cite{7445130,9026753,9219133,9110865}, it may be unnecessary to use the DS-FPS architecture for enhancing energy efficiency at the cost of spectral efficiency degradation. Hence, the DS-FPS architecture is THz-specific.

We set the number of RF chains as $L_t$ and each RF chain connects with $Q$ FPSs. The provided phases of the FPSs are fixed as $\Phi_1$, $\Phi_2$, ..., $\Phi_Q$, respectively. One drawback of FPS is that the provided phase is fixed while the required phase to steer beams varies with different channels, which thereby compromises the spectral efficiency performance. To overcome this drawback, we propose a switch network to enable dynamic connection, where each antenna can select one FPS from all $L_tQ$ FPSs to connect with, i.e., dynamically selects one proper phase from $\Phi_1$, $\Phi_2$, ..., $\Phi_Q$ to perform analog beamforming. Since the required phase of analog beamforming at each antenna may be an arbitrary value in $[0,2\pi]$ when channel varies, we set the phases of FPSs uniformly located in $[0,2\pi]$, i.e., $\Phi_{i}=\frac{2\pi(i-1)}{Q}$ for $i=1,2,\ldots,Q$, to make the beamforming weight error at each antenna no larger than $\frac{\pi}{Q}$.
The system model of the DS-FPS hybrid beamforming architecture at the $k^{\rm th}$ sub-carrier can be expressed as
\begin{equation}
\textbf{y}[k] = \textbf{W}[k]^H\textbf{H}[k]\textbf{S}\textbf{F}\textbf{D}[k]\textbf{s}[k] +  \textbf{W}[k]^H\textbf{n}[k],
\label{system_model_DS}
\end{equation}
where $\textbf{s}[k]\in\mathbb{C}^{N_s\times1}$ and $\textbf{y}[k]\in\mathbb{C}^{N_s\times1}$ denote the transmitted and received signals. $N_s$ is number of data streams. $\textbf{n}[k]\in\mathbb{C}^{N_r\times 1}$ is the noise vector. $\textbf{W}[k]\in\mathbb{C}^{N_r\times N_s}$ is the combining matrix at receiver. $N_t\times L_tQ$-dimensional binary matrix $\textbf{S}$ and $L_tQ\times L_t$-dimensional matrix $\textbf{F}$ represent the switch network matrix and the phase matrix of the FPS network, respectively. The phase of one FPS is the same for each sub-carrier. The state of switch is identical for each sub-carrier. Hence, the frequency index $[k]$ is omitted in $\textbf{F}$ and $\textbf{S}$. As a result, $\textbf{F}$ can be written as 
\begin{equation}
\textbf{F}={\rm blkdiag}(\underbrace{\textbf{f},\ldots,\textbf{f}}_{L_t}),
\label{structure_FPS_network}
\end{equation}
where $\textbf{f}=[e^{j\Phi_1},e^{j\Phi_2},\ldots,e^{j\Phi_Q}]^T$ represents the phase vector generated by $Q$ FPSs of each RF chain. Since each antenna only selects one FPS from all $L_tQ$ FPSs through one closed switch, each row of $\textbf{S}$ has only one `1' and other elements are `0's, i.e., $\lVert\textbf{S}_i\rVert_{0}=1, i=1,2,\ldots,N_t$, where $\textbf{S}_{i}$ denotes the $i^{\rm th}$ row of $\textbf{S}$.
$\textbf{D}[k]\in\mathbb{C}^{L_t\times N_s}$ represents the digital beamforming matrix. The transmit power constraint is enforced on $\textbf{D}[k]$ as $\sum_{k=1}^{K}\lVert\textbf{S}\textbf{F}\textbf{D}[k]\rVert^{2}_{F}=\rho$, where $\rho$ is the transmit power of the DS-FPS architecture.

\subsection{Design Problem: Maximize the Spectral Efficiency with Partial CSI}
The spectral efficiency of the DS-FPS architecture can be expressed as~\cite{7389996}
\begin{align}
SE\!=\!\frac{1}{K}\sum\nolimits_{k=1}^{K}\!\!\!\!{\rm{log}}_2\Big(\Big\lvert\textbf{I}_{N_r}\!+\!\frac{1}{\sigma^{2}_{k}}\textbf{W}[k](\textbf{W}[k]^H\textbf{W}[k])^{-1}\textbf{W}[k]^H\textbf{H}[k]\textbf{S}\textbf{F}\textbf{D}[k]\textbf{D}[k]^H\textbf{F}^H\textbf{S}^H\textbf{H}[k]^{H}\Big\rvert\Big),
\label{SE_formulation}
\end{align}
where $\sigma^{2}_{k}$ denotes the noise power of the $k^{\rm th}$ sub-carrier. 
To focus on the analysis of the DS-FPS architecture at the transmitter, we consider that the receiver is arranged with optimal digital combining, i.e., $\textbf{W}[k]$ is the first $N_s$ columns of the left singular matrix of $\textbf{H}[k]$. It has been studied in~\cite{7389996,7445130} that, when designing the transmitter to maximize the spectral efficiency, the influence of the combining matrix $\textbf{W}[k]$ at the receiver can be decoupled. Hence, the design problem of $\textbf{S}$ and $\textbf{D}[k]$ can be formulated as~\cite{7389996,7445130}
\begin{subequations}
	\begin{align}
	&\mathop{\rm max\ }\limits_{\textbf{S},\textbf{D}[k]}\frac{1}{K}\sum\nolimits_{k=1}^{K}{\rm{log}}_2\Big(\Big\lvert\textbf{I}_{N_r}+\frac{1}{\sigma^{2}_{k}}\textbf{H}[k]\textbf{S}\textbf{F}\textbf{D}[k]\textbf{D}[k]^H\textbf{F}^H\textbf{S}^H\textbf{H}[k]^{H}\Big\rvert\Big)
	\label{design_problem_objective}\\ 
	&\mathrm{s.t.}\ \textbf{S}_{i,l}\in\{0,1\}, \lVert\textbf{S}_{i}\rVert_{0}=1,\forall i,l
	\label{design_problem_constraints_1}\\ &\quad \ \ \sum\nolimits_{k=1}^{K}\lVert\textbf{S}\textbf{F}\textbf{D}[k]\rVert^{2}_{F}=\rho, 
	\label{design_problem_constraints_2}
	\end{align}
	\label{design_problem}%
\end{subequations}
where $\textbf{S}_{i,l}$ denotes the element at the $i^{\rm th}$ row and the $l^{\rm th}$ column of $\textbf{S}$. 
In this work, we focus on the solution to~\eqref{design_problem} at the transmitter, while the design of the DS-FPS architecture at the receiver side is similar, by rewriting~\eqref{design_problem} into the form of combining matrices and applying the proposed algorithms in the following.

\section{Row-successive-decomposition (RSD) Algorithm}
\label{section_RSD_algorithm}
In this section, we propose an RSD algorithm to solve the design problem. 
One main challenge is that the objective function is related to the full CSI $\textbf{H}[k]$, while the transmitter only knows $\textbf{A}_t[k]$ and $\bm{\bar\Lambda}[k]$. To tackle this problem, we first derive an approximated form of the objective function~\eqref{design_problem_objective}, which only relies on $\textbf{A}_t[k]$ and $\bm{\bar\Lambda}[k]$ and excludes the unknown $\textbf{H}[k]$. Then, we decompose the rows of switch network matrix $\textbf{S}$ successively to transform the intractable design problem as multiple tractable sub-problems, to overcome the non-convex binary constraint of $\textbf{S}$.

\subsection{Approximated Form of~\eqref{design_problem_objective}}
As analyzed in~\eqref{channel_model_planar}, $\textbf{H}[k]=\textbf{A}_{r}[k]\bm{\Lambda}[k]\textbf{A}_{t}[k]^H=\textbf{A}_{r}[k]\left(\bm{\bar\Lambda}[k]\odot e^{j*\bm{\dot\Lambda}[k]}\right)\textbf{A}_{t}[k]^H$. Hence, the objective function~\eqref{design_problem_objective} can be rearranged as
\begin{subequations}
\begin{align}
&\quad \ \frac{1}{K}\sum\nolimits_{k=1}^{K}{\rm{log}}_2\Big(\Big\lvert\textbf{I}_{N_r}+(1/\sigma_k^2)\textbf{H}[k]\textbf{S}\textbf{F}\textbf{D}[k]\textbf{D}[k]^H\textbf{F}^H\textbf{S}^H\textbf{H}[k]^{H}\Big\rvert\Big)\\
&=\frac{1}{K}\sum\nolimits_{k=1}^{K}{\rm{log}}_2\Big(\Big\lvert\textbf{I}_{N_r}+(1/\sigma_k^2)\textbf{A}_{r}[k]\bm{\Lambda}[k]\textbf{A}_{t}[k]^H\textbf{G}[k]\textbf{A}_{t}[k]\bm{\Lambda}[k]^H\textbf{A}_{r}[k]^H\Big\rvert\Big)\label{expression_SE_approx_1}\\
&=\frac{1}{K}\sum\nolimits_{k=1}^{K}{\rm{log}}_2\Big(\Big\lvert\textbf{I}_{N_p}+(1/\sigma_k^2)\bm{\Lambda}[k]^H\textbf{A}_{r}[k]^H\textbf{A}_{r}[k]\bm{\Lambda}[k]\textbf{A}_{t}[k]^H\textbf{G}[k]\textbf{A}_{t}[k]\Big\rvert\Big)\label{expression_SE_approx_2}\\
&\approx\frac{1}{K}\sum\nolimits_{k=1}^{K}{\rm{log}}_2\Big(\Big\lvert\textbf{I}_{N_p}+(N_r/\sigma_k^2)\bm{\bar\Lambda}[k]^2\textbf{A}_{t}[k]^H\textbf{G}[k]\textbf{A}_{t}[k]\Big\rvert\Big)\label{expression_SE_approx_3}\\
&=\frac{1}{K}\sum\nolimits_{k=1}^{K}{\rm{log}}_2\Big(\Big\lvert\textbf{I}_{N_p}+(N_r/\sigma_k^2)\bm{\bar\Lambda}[k]\textbf{A}_{t}[k]^H\textbf{S}\textbf{F}\textbf{D}[k]\textbf{D}[k]^H\textbf{F}^H\textbf{S}^H\textbf{A}_{t}[k]\bm{\bar\Lambda}[k]^H\Big\rvert\Big),
\label{expression_SE_approx_4}
\end{align}
\end{subequations}
where $\textbf{G}[k]=\textbf{S}\textbf{F}\textbf{D}[k]\textbf{D}[k]^H\textbf{F}^H\textbf{S}^H$ is included in~\eqref{expression_SE_approx_1}.
Moreover, \eqref{expression_SE_approx_2} comes from the property of the determinant that ${\rm log}_2(\lvert\textbf{I}+\textbf{X}\textbf{Y}\rvert)={\rm log}_2(\lvert\textbf{I}+\textbf{Y}\textbf{X}\rvert)$, where $\textbf{X}=\textbf{A}_{r}[k]\bm{\Lambda}[k]\textbf{A}_{t}[k]^H\textbf{G}[k]\textbf{A}_{t}[k]$ and $\textbf{Y}=\bm{\Lambda}[k]^H\textbf{A}_{r}[k]^H$.
In addition, \eqref{expression_SE_approx_3} follows the \textit{Approximation 1} as below, i.e., $\textbf{A}_r[k]^H\textbf{A}_r[k]\approx N_r\textbf{I}_{N_p}$. Consequently, we have $\bm{\Lambda}[k]^H\textbf{A}_{r}[k]^H\textbf{A}_{r}[k]\bm{\Lambda}[k]\approx N_r\bm{\Lambda}[k]^H\bm{\Lambda}[k]$. Moreover, since $\bm{\Lambda}[k]$ is a diagonal and square matrix as analyzed in~\eqref{channel_model_planar}, we have $\bm{\Lambda}[k]^H\bm{\Lambda}[k]=\bm{\bar\Lambda}[k]^2$. Last, \eqref{expression_SE_approx_4} follows the property of the determinant similar to~\eqref{expression_SE_approx_2}.

\textit{Approximation 1:} In THz UM-MIMO systems, $\textbf{A}_r[k]^H\textbf{A}_r[k]\approx N_r\textbf{I}_{N_p}$, for $k=1$, $2$, ..., $K$.

\textit{Proof:} The $i^{\rm th}$ column of $\textbf{A}_{r}[k]$ is $\textbf{a}_{ri}[k]$ in~\eqref{steering_vector_UPA}. The element of $\textbf{A}_{r}[k]^H\textbf{A}_{r}[k]$ at the $i^{\rm th}$ row and the $l^{\rm th}$ column can be represented as $\textbf{a}_{ri}[k]^H\textbf{a}_{rl}[k]$.
Hence, \textit{Approximation 1} is equivalent to
\begin{equation}
\frac{1}{N_r}\textbf{a}_{ri}[k]^H\textbf{a}_{rl}[k]\approx\mathbbm{1}(i=l),
\label{equivalent_approx_1}	
\end{equation}
where $\mathbbm{1}(i=l)$ is the indicator function that $\mathbbm{1}=1$ when $i=l$ and $\mathbbm{1}=0$ when $i\neq l$.
According to the structure of $\textbf{a}_{ri}[k]$ in~\eqref{steering_vector_UPA}, when $i=l$, $\frac{1}{N_r}\textbf{a}_{ri}[k]^H\textbf{a}_{rl}[k]$ is indeed equal to $1$. Next, we show that $\frac{1}{N_r}\lvert\textbf{a}_{ri}[k]^H\textbf{a}_{rl}[k]\rvert\approx0$ when $i\neq l$, which leads to $\frac{1}{N_r}\textbf{a}_{ri}[k]^H\textbf{a}_{rl}[k]\approx0$. Specifically, $\frac{1}{N_r}\lvert\textbf{a}_{ri}[k]^H\textbf{a}_{rl}[k]\rvert$ can be expressed as
\begin{subequations}
\begin{align}
&\qquad\qquad\quad \ \begin{aligned}\frac{1}{N_r}\Big\lvert\Big(\big[1&,
... ,e^{-j\frac{2\pi}{\lambda_k}d(L-1){\rm sin}(\phi_{ri}){\rm sin}(\theta_{ri})}\big]\otimes\big[1,
... ,e^{-j\frac{2\pi}{\lambda_k}d(W-1){\rm cos}(\theta_{ri})}\big]\Big)\\&\times\Big(\big[1,
... ,e^{j\frac{2\pi}{\lambda_k}d(L-1){\rm sin}(\phi_{rl}){\rm sin}(\theta_{rl})}\big]^T\otimes\big[1,
... ,e^{j\frac{2\pi}{\lambda_k}d(W-1){\rm cos}(\theta_{rl})}\big]^T\Big)\Big\lvert
\end{aligned}
\label{approx_leq_1}\\
&\qquad\qquad\ \! =\frac{1}{N_r}\left\lvert\sum\nolimits_{a=0}^{L-1} e^{j\frac{2\pi d}{\lambda_k}a({\rm sin}(\phi_{rl}){\rm sin}(\theta_{rl})-{\rm sin}(\phi_{ri}){\rm sin}(\theta_{ri}))}\sum\nolimits_{b=0}^{W-1} e^{j\frac{2\pi d}{\lambda_k}b({\rm cos}(\theta_{rl})-{\rm cos}(\theta_{ri}))}\right\rvert
\label{approx_leq_2}\\
&\qquad\qquad\ \!=\frac{1}{N_r}\left\lvert\frac{{\rm sin}(\frac{\pi d}{\lambda_k}L\Psi_1)}{{\rm sin}(\frac{\pi d}{\lambda_k}\Psi_1)}\times\frac{{\rm sin}(\frac{\pi d}{\lambda_k}W\Psi_2)}{{\rm sin}(\frac{\pi d}{\lambda_k}\Psi_2)}\right\rvert
\label{approx_leq_3}\\
&\qquad\qquad\ \!\leq\frac{1}{N_r}\frac{1}{\big\lvert{\rm sin}(\frac{\pi d}{\lambda_k}\Psi_1){\rm sin}(\frac{\pi d}{\lambda_k}\Psi_2)\big\rvert},
\label{approx_leq_4}
\end{align}
\end{subequations}
where~\eqref{approx_leq_2} follows the mixed-product property of the Kronecker product. 
$\Psi_1={\rm sin}(\phi_{rl}){\rm sin}(\theta_{rl})-{\rm sin}(\phi_{ri}){\rm sin}(\theta_{ri})$, $\Psi_2={\rm cos}(\theta_{rl})-{\rm cos}(\theta_{ri})$ in~\eqref{approx_leq_3}. \eqref{approx_leq_4} comes from the fact that ${\rm sin}(\frac{\pi d}{\lambda_k}L\Psi_1)\leq1$ and ${\rm sin}(\frac{\pi d}{\lambda_k}W\Psi_2)\leq1$. Since $i\neq l$, we have $\Psi_1\neq0$, $\Psi_2\neq0$, and $\lvert{\rm sin}(\frac{\pi d}{\lambda_k}\Psi_1){\rm sin}(\frac{\pi d}{\lambda_k}\Psi_2)\rvert\neq0$. Therefore, in THz UM-MIMO systems with ultra-massive antennas, e.g., $N_r\geq1024$, $\frac{1}{N_r}\lvert\textbf{a}_{ri}^H\textbf{a}_{rl}\rvert\leq\frac{1}{N_r}\frac{1}{\lvert{\rm sin}(\frac{\pi d}{\lambda_k}\Psi_1){\rm sin}(\frac{\pi d}{\lambda_k}\Psi_2)\rvert}$ is usually very close to $0$ and the approximation $\frac{1}{N_r}\lvert\textbf{a}_{ri}[k]^H\textbf{a}_{rl}[k]\rvert\approx0$ holds. The special case is that when the directions of the $i^{\rm th}$ path and the $l^{\rm th}$ path are similar, $\Psi_1$ and $\Psi_2$ are small such that $\lvert{\rm sin}(\frac{\pi d}{\lambda_k}\Psi_1){\rm sin}(\frac{\pi d}{\lambda_k}\Psi_2)\rvert$ is close to $0$ and the approximation error of $\frac{1}{N_r}\lvert\textbf{a}_{ri}[k]^H\textbf{a}_{rl}[k]\rvert\approx0$ is large. 

\begin{figure*}
	\setlength{\belowcaptionskip}{0pt}
	\centering
	\captionsetup{font={footnotesize}}
	\subfigure[Formula~\eqref{approx_leq_1} versus $\theta_{rl}$ and $\phi_{rl}$.]{
		\includegraphics[scale=0.26]{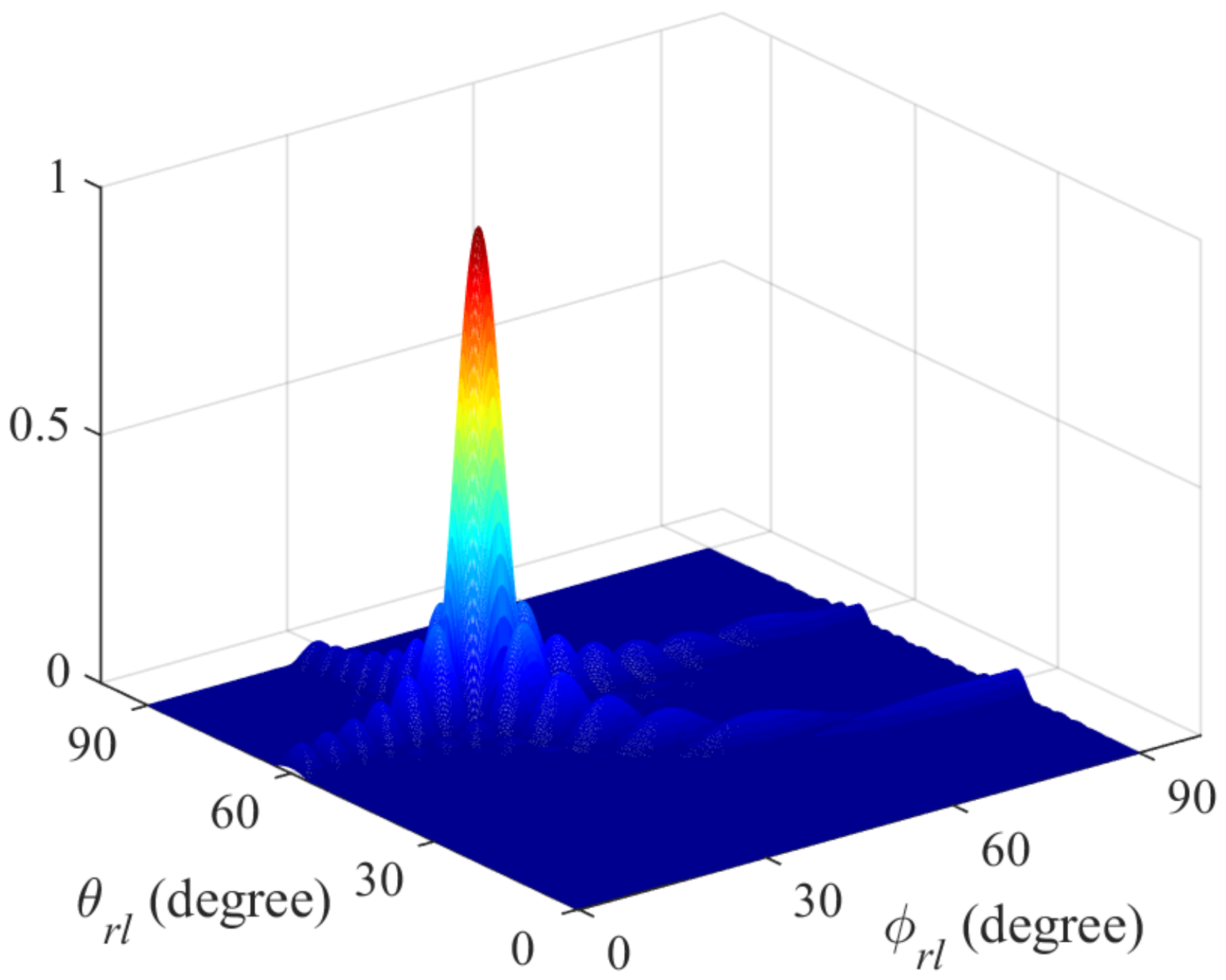}}
	\subfigure[Vertical view of Fig.~\ref{fig_approx}(a).]{
		\includegraphics[scale=0.26]{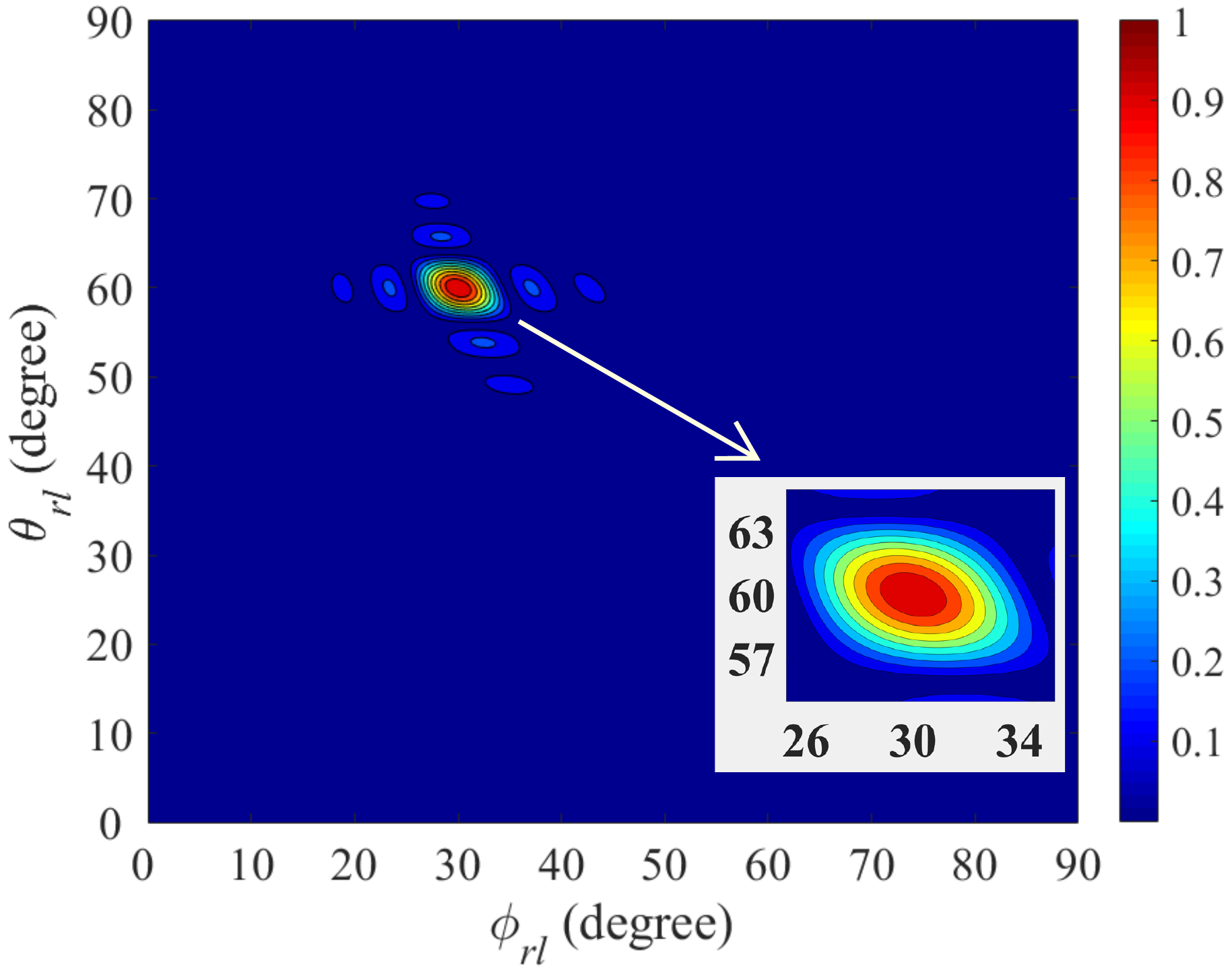}}
	\caption{$\frac{1}{N_r}\lvert\textbf{a}_{ri}[k]^H\textbf{a}_{rl}[k]\rvert$ in~\eqref{approx_leq_1} versus $\theta_{rl}$ and $\phi_{rl}$, where $\theta_{ri}=60^{\circ}$ and $\phi_{ri}=30^{\circ}$. $L=W=32$, $N_r=L\times W=1024$, $\lambda_k=1$ mm ($0.3$ THz), $d=0.5\lambda_k$.}
	\label{fig_approx}
	\vspace{-7.5mm}
\end{figure*}
We further assess the approximation error in Fig.~\ref{fig_approx}. The direction of the $i^{\rm th}$ path is $\phi_{ri}=30^{\circ}$ and $\theta_{ri}=60^{\circ}$. As shown in Fig.~\ref{fig_approx}(a) and Fig.~\ref{fig_approx}(b), we plot $\frac{1}{N_r}\lvert\textbf{a}_{ri}[k]^H\textbf{a}_{rl}[k]\rvert$ versus the direction of the $l^{\rm th}$ path, i.e., $\phi_{rl}$ and $\theta_{rl}$. Due to the symmetrical property of the sinusoidal function, we only consider the cases that $0^{\circ}\leq\phi_{rl}\leq90^{\circ}$ and $0^{\circ}\leq\theta_{rl}\leq90^{\circ}$, while the remaining cases are similar and extensible. For most angles of $\phi_{rl}$ and $\theta_{rl}$ which locate in the blue region, $\frac{1}{N_r}\lvert\textbf{a}_{ri}[k]^H\textbf{a}_{rl}[k]\rvert\approx0$. When $\phi_{rl}$ and $\theta_{rl}$ are very close to $\phi_{ri}$ and $\theta_{ri}$, i.e., the $l^{\rm th}$ path locates in the highlighted region $\{(\phi_{rl},\theta_{rl})\big\lvert\phi_{ri}-4^{\circ}<\phi_{rl}<\phi_{ri}+4^{\circ}\cap\theta_{ri}-3^{\circ}<\theta_{rl}<\theta_{ri}+3^{\circ}\}$, $\frac{1}{N_r}\lvert\textbf{a}_{ri}[k]^H\textbf{a}_{rl}[k]\rvert$ is not close to $0$. The blue region accounts for a very large proportion of the space, i.e., more than $99\%$. Moreover, the THz channel is usually sparse, e.g., the number of multipath in the spatial domain is around $5$~\cite{6998944}. In light of these, the probability that the $l^{\rm th}$ path locates in the highlighted region is very small.
Without loss of generality, we can state that the approximation $\frac{1}{N_r}\lvert\textbf{a}_{ri}[k]^H\textbf{a}_{rl}[k]\rvert\approx0$, i.e., $\frac{1}{N_r}\textbf{a}_{ri}[k]^H\textbf{a}_{rl}[k]\approx0$, holds reasonably well.
\hfill $\blacksquare$

\subsection{Design of Digital Beamforming Matrix {\rm \textbf{D}[{\it k}]}}
Till now, we obtain an approximated form~\eqref{expression_SE_approx_4} of the original objective function~\eqref{design_problem_objective}, which relies on the partial CSI $\bm{\bar\Lambda}[k]$ and $\textbf{A}_t[k]^H$. Next, we present how to use~\eqref{expression_SE_approx_4} to design $\textbf{D}[k]$ and $\textbf{S}$.
We first analyze the solution of $\textbf{D}[k]$ by assuming that $\textbf{S}$ has been determined. 
The maximization of the original objective~\eqref{design_problem_objective} can be transformed as the maximization of~\eqref{expression_SE_approx_4} as
\begin{equation}
	\mathop{\rm max\ }\limits_{\textbf{S},\textbf{D}[k]}\frac{1}{K}\sum\nolimits_{k=1}^{K}{\rm{log}}_2\Big(\Big\lvert\textbf{I}_{N_p}+\frac{N_r}{\sigma^{2}_{k}}\bm{\bar\Lambda}[k]\textbf{A}_{t}[k]^H\textbf{S}\textbf{F}\textbf{D}[k]\textbf{D}[k]^H\textbf{F}^H\textbf{S}^H\textbf{A}_{t}[k]\bm{\bar\Lambda}[k]^H\Big\rvert\Big),
	\label{design_problem_1}
\end{equation}
where the constraints of $\textbf{S}$ and $\textbf{D}[k]$ are the same as in~\eqref{design_problem}.
Assuming that $\textbf{S}$ has been determined and omitting the transmit power constraint temporarily, the solution of $\textbf{D}[k]$ to maximize \eqref{design_problem_1} is
\begin{equation}
	\textbf{D}[k]=\widetilde{\textbf{V}}_{N_s}[k]\widetilde{\bm{\Gamma}}[k],
	\label{solution_D_RSD}
\end{equation}
where $\widetilde{\textbf{V}}_{N_s}[k]$ is the first $N_s$ columns of $\widetilde{\textbf{V}}[k]$, which comes from the singular value decomposition (SVD) of $\textbf{H}_e[k]=\bm{\bar\Lambda}[k]\textbf{A}_t[k]^H\textbf{S}\textbf{F}$ given by $\textbf{H}_e[k]=\widetilde{\textbf{U}}[k]\widetilde{\bm{\Sigma}}[k]\widetilde{\textbf{V}}[k]^H$.
Moreover, $\widetilde{\bm{\Gamma}}[k]$ is the power allocation matrix, for which the water-filling allocation is the optimal. Despite so, to reduce the computational complexity, we consider the more practical equal-power allocation such that $\widetilde{\bm{\Gamma}}[k]=\textbf{I}_{N_s}$. The consideration of water-filling power allocation and the corresponding analog and digital beamforming solution can be considered in the future work.
\subsection{Design of Switch Network Matrix {\rm \textbf{S}}}
Next, we successively decompose the rows of $\textbf{S}$ to enable the design of $\textbf{S}$ for the maximization of~\eqref{design_problem_1}.
By substituting the solution of $\textbf{D}[k]$ in \eqref{solution_D_RSD} into \eqref{design_problem_1}, the $k^{\rm th}$ term of~\eqref{design_problem_1} is derived as
\begin{subequations}
	\begin{align}
	&\quad \ {\rm{log}}_2\Big(\Big\lvert\textbf{I}_{N_p}+(N_r/\sigma_k^2)\bm{\bar\Lambda}[k]\textbf{A}_{t}[k]^H\textbf{S}\textbf{F}\textbf{D}[k]\textbf{D}[k]^H\textbf{F}^H\textbf{S}^H\textbf{A}_{t}[k]\bm{\bar\Lambda}[k]^H\Big\rvert\Big)
	\label{relaxation_SE_1}\\
	&={\rm{log}}_2\Big(\Big\lvert\textbf{I}_{N_p}+(N_r/\sigma_k^2)\textbf{H}_e[k]\widetilde{\textbf{V}}_{N_s}\textbf{I}_{N_s}\textbf{I}_{N_s}^H\widetilde{\textbf{V}}_{N_s}^H\textbf{H}_e[k]^H\Big\rvert\Big)
	\label{relaxation_SE_2}\\
	&={\rm{log}}_2\Big(\Big\lvert\textbf{I}_{N_p}+(N_r/\sigma_k^2)\widetilde{\textbf{U}}\widetilde{\bm{\Sigma}}\widetilde{\textbf{V}}^H\widetilde{\textbf{V}}_{N_s}\widetilde{\textbf{V}}_{N_s}^H\widetilde{\textbf{V}}\widetilde{\bm{\Sigma}}^H\widetilde{\textbf{U}}^H\Big\rvert\Big)
	\label{relaxation_SE_3}\\
	&={\rm{log}}_2\Big(\Big\lvert\textbf{I}_{N_p}+(N_r/\sigma_k^2)\widetilde{\bm{\Sigma}}^H\widetilde{\textbf{U}}^H\widetilde{\textbf{U}}\widetilde{\bm{\Sigma}}[\widetilde{\textbf{V}}_{N_s},\widetilde{\textbf{V}}_{\epsilon} ]^H\widetilde{\textbf{V}}_{N_s}\widetilde{\textbf{V}}_{N_s}^H[\widetilde{\textbf{V}}_{N_s},\widetilde{\textbf{V}}_{\epsilon}]\Big\rvert\Big)
    \label{relaxation_SE_4}\\
	&={\rm{log}}_2\Big(\Big\lvert\textbf{I}_{N_s}+(N_r/\sigma_k^2)\widetilde{\bm{\Sigma}}_{N_s}^2\Big\rvert\Big)
    \label{relaxation_SE_5}\\    
	&\leq{\rm{log}}_2\Big(\Big\lvert\textbf{I}_{L_t}+(N_r/\sigma_k^2)\widetilde{\bm{\Sigma}}_{L_t}^2\Big\rvert\Big)
	\label{relaxation_SE_6}\\
	&={\rm{log}}_2\Big(\Big\lvert\textbf{I}_{N_p}+(N_r/\sigma_k^2)\bm{\bar\Lambda}[k]\textbf{A}_{t}[k]^H\textbf{S}\textbf{F}\textbf{F}^H\textbf{S}^H\textbf{A}_{t}[k]\bm{\bar\Lambda}[k]^H\Big\rvert\Big).
    \label{relaxation_SE_8}	
	\end{align}
\end{subequations}
From~\eqref{relaxation_SE_2} to~\eqref{relaxation_SE_8}, for $\widetilde{\textbf{U}}[k]$, $\widetilde{\bm{\Sigma}}[k]$, $\widetilde{\textbf{V}}[k]$, and $\widetilde{\textbf{V}}_{N_s}[k]$, we omit the index $[k]$ for simplicity.
\eqref{relaxation_SE_3} is the result of applying the SVD of $\textbf{H}_e[k]$ given by $\textbf{H}_e[k]=\widetilde{\textbf{U}}\widetilde{\bm{\Sigma}}\widetilde{\textbf{V}}^H$. \eqref{relaxation_SE_4} comes from the property that ${\rm log}_2(\lvert\textbf{I}+\textbf{X}\textbf{Y}\rvert)={\rm log}_2(\lvert\textbf{I}+\textbf{Y}\textbf{X}\rvert)$, where $\textbf{X}=\widetilde{\textbf{U}}\widetilde{\bm{\Sigma}}\widetilde{\textbf{V}}^H\widetilde{\textbf{V}}_{N_s}\widetilde{\textbf{V}}_{N_s}^H\widetilde{\textbf{V}}$ and $\textbf{Y}=\widetilde{\bm{\Sigma}}^H\widetilde{\textbf{U}}^H$. Moreover, $\widetilde{\textbf{V}}_{N_s}$ and $\widetilde{\textbf{V}}_{\epsilon}$ denote the first $N_s$ columns and the remaining columns of $\widetilde{\textbf{V}}$, respectively. \eqref{relaxation_SE_5} follows the property of SVD that $\widetilde{\textbf{U}}^H\widetilde{\textbf{U}}=\textbf{I}$, $\widetilde{\textbf{V}}_{N_s}^H\widetilde{\textbf{V}}_{N_s}=\textbf{I}$, and $\widetilde{\textbf{V}}_{\epsilon}^H\widetilde{\textbf{V}}_{N_s}=\textbf{0}$, where $\widetilde{\bm{\Sigma}}_{N_s}$ represents the first $N_s$ rows and $N_s$ columns of $\widetilde{\bm{\Sigma}}$. In the hybrid beamforming system, the number of RF chains $L_t$ is usually slightly larger than or equal to the number of data streams $N_s$, since the spectral efficiency enhancement brought by additional RF chains is negligible~\cite{7397861}. Therefore, we use \eqref{relaxation_SE_6} as an upper bound on \eqref{relaxation_SE_5}, where $\widetilde{\bm{\Sigma}}_{L_t}$ represents the first $L_t$ rows and $L_t$ columns of $\widetilde{\bm{\Sigma}}$ and the equality holds when $N_s=L_t$. The dimension of $\textbf{H}_e[k]$ is $N_p\times L_t$ such that $\textbf{H}_e[k]$ has at most $L_t$ non-zero singular values, which suggests that $\widetilde{\bm{\Sigma}}_{L_t}$ contains all the non-zero singular values of $\textbf{H}_e[k]$. According to the property of SVD, we obtain ${\rm{log}}_2(\lvert\textbf{I}_{L_t}+(N_r/\sigma_k^2)\widetilde{\bm{\Sigma}}_{L_t}^2\rvert)={\rm{log}}_2(\lvert\textbf{I}_{N_p}+(N_r/\sigma_k^2)\textbf{H}_e[k]\textbf{H}_e[k]^H\rvert)$, which equals to \eqref{relaxation_SE_8}.

So far, we have derived an upper bound according to \eqref{relaxation_SE_8} for~\eqref{relaxation_SE_1}, which is uncorrelated with $\textbf{D}[k]$ such that the coupling between $\textbf{D}[k]$ and $\textbf{S}$ is mitigated. Next, we propose to design $\textbf{S}$ to maximize \eqref{relaxation_SE_8} rather than directly maximizing \eqref{relaxation_SE_1}.
One main difficulty to solve $\textbf{S}$ is the binary constraint~\eqref{design_problem_constraints_1} on each row. 
The authors in~\cite{7445130} proposed to successively decompose the hybrid beamforming matrix in the column-manner to enable the design. Inspired by this, to tackle the row-manner constraint of $\textbf{S}$, we propose to decompose each row of $\textbf{S}$ successively. Although the idea of successive decomposition is similar, the detailed constraints and derivations of the row-manner decomposition in this work are quite different from the column-manner decomposition in~\cite{7445130}. Moreover, some important elements of the RSD algorithm, including the derived approximated spectral efficiency~\eqref{expression_SE_approx_4}, the solution of $\textbf{D}[k]$ as well as the derived upper bound~\eqref{relaxation_SE_8} on the approximated spectral efficiency, and the solution of each row of $\textbf{S}$ have not been studied in~\cite{7445130}.
To start with, \eqref{relaxation_SE_8} can be rewritten as 
\begin{subequations}
	\begin{align}
	&\quad \
	{\rm{log}}_2\Big(\Big\lvert\textbf{I}_{N_p}\!+\!(N_r/\sigma_k^2)\bm{\bar\Lambda}[k]\textbf{A}_{t}[k]^H\textbf{S}\textbf{F}\textbf{F}^H\textbf{S}^H\textbf{A}_{t}[k]\bm{\bar\Lambda}[k]^H\Big\rvert\Big)
	\label{RSD_SIC_1}\\
	&= {\rm{log}}_2\Big(\Big\lvert\textbf{I}_{N_p}\!+\!(N_r/\sigma_k^2)\bm{\bar\Lambda}[k]^2\textbf{A}_{t}[k]^H\textbf{S}\textbf{F}\textbf{F}^H\textbf{S}^H\textbf{A}_{t}[k]\Big\rvert\Big)
	\label{RSD_SIC_2}\\
	&={\rm{log}}_2\bigg(\bigg\lvert\textbf{I}_{N_p}\!+\!\frac{N_r}{\sigma^{2}_{k}}\bm{\bar\Lambda}[k]^2\!\Big[\textbf{C}_{1:N_t-1}[k],\textbf{C}_{N_t}[k]\Big]\!\!\bigg[\!\!\begin{array}{c}\textbf{S}_{1:N_t-1}\\\textbf{S}_{N_t}\end{array}\!\!\!\bigg]\!\textbf{F}\textbf{F}^H\![\textbf{S}_{1:N_t-1}^H,\textbf{S}_{1:N_t}^H]\!\bigg[\!\!\begin{array}{c}\textbf{C}_{1:N_t-1}[k]^H\\\textbf{C}_{N_t}[k]^H\end{array}\!\!\bigg]\bigg\rvert\bigg)
	\label{RSD_SIC_3}\\	
	&={\rm{log}}_2\Big(\Big\lvert\textbf{I}_{N_p}\!+\!(N_r/\sigma_k^2)\bm{\bar\Lambda}[k]^2(\textbf{A}[k]\textbf{A}[k]^H\!+\!\textbf{B}[k]\textbf{A}[k]^H\!+\!\textbf{A}[k]\textbf{B}[k]^H\!+\!\textbf{B}[k]\textbf{B}[k]^H\!)\Big\rvert\Big)
	\label{RSD_SIC_4}\\
	&\ \begin{aligned}	
	={\rm{log}}_2\Big(\Big\lvert\textbf{I}_{N_p}&\!+(N_r/\sigma_k^2)\bm{\bar\Lambda}[k]^2\textbf{A}[k]\textbf{A}[k]^H\Big\rvert\Big)\\&+{\rm{log}}_2\Big(\Big\lvert\textbf{I}_{N_p}\!+\!(N_r/\sigma_k^2)\textbf{T}[k]^{-1}\bm{\bar\Lambda}[k]^2(\textbf{B}[k]\textbf{A}[k]^H\!+\!\textbf{A}[k]\textbf{B}[k]^H\!+\!\textbf{B}[k]\textbf{B}[k]^H\!\Big\rvert\Big).
	\end{aligned}
	\label{RSD_SIC_5}
	\end{align}
\end{subequations}
In \eqref{RSD_SIC_3}, we use $\textbf{C}[k]$ to represent $\textbf{A}_t^H[k]$, where $\textbf{C}_{1:p}[k]$ and $\textbf{C}_{q}[k]$ denote the first $p$ columns and the $q^{\rm th}$ column of $\textbf{C}[k]$, respectively. $\textbf{S}_{1:p}$ and $\textbf{S}_{q}$ denote the first $p$ rows and the $q^{\rm th}$ row of $\textbf{S}$, respectively. In \eqref{RSD_SIC_4}, $\textbf{A}[k]=\textbf{C}_{1:N_t-1}[k]\textbf{S}_{1:N_t-1}\textbf{F}$ and $\textbf{B}[k]=\textbf{C}_{N_t}[k]\textbf{S}_{N_t}\textbf{F}$. \eqref{RSD_SIC_5} comes from the property of determinant that $\lvert\textbf{I}+\textbf{X}+\textbf{Y}\rvert=\lvert\textbf{I}+\textbf{X}\rvert\cdot\lvert\textbf{I}+(\textbf{I}+\textbf{X})^{-1}\textbf{Y}\rvert$, where $\textbf{X}=\frac{N_r}{\sigma^{2}_{k}}\bm{\bar\Lambda}[k]^2\textbf{A}[k]\textbf{A}[k]^H$, $\textbf{Y}=\frac{N_r}{\sigma^{2}_{k}}\bm{\bar\Lambda}[k]^2(\textbf{B}[k]\textbf{A}[k]^H+\textbf{A}[k]\textbf{B}[k]^H+\textbf{B}[k]\textbf{B}[k]^H)$, and $\textbf{T}[k]=\textbf{I}_{N_p}+\frac{N_r}{\sigma^{2}_{k}}\bm{\bar\Lambda}[k]^2\textbf{A}[k]\textbf{A}[k]^H$.

We observe that, by substituting $\textbf{A}[k]=\textbf{C}_{1:N_t-1}[k]\textbf{S}_{1:N_t-1}\textbf{F}$, the first term in~\eqref{RSD_SIC_5} can be further expressed as ${\rm{log}}_2\Big(\Big\lvert\textbf{I}_{N_p}+\frac{N_r}{\sigma^{2}_{k}}\bm{\bar\Lambda}[k]^2\textbf{C}_{1:N_t-1}[k]\textbf{S}_{1:N_t-1}\textbf{F}\textbf{F}^H\textbf{S}_{1:N_t-1}^H\textbf{C}_{1:N_t-1}[k]^H\Big\rvert\Big)$, which has the similar structure with \eqref{RSD_SIC_2}. Therefore, we can continue to decompose the first term of~\eqref{RSD_SIC_5} with the similar procedures from~\eqref{RSD_SIC_2} to~\eqref{RSD_SIC_5}. After $N_t$ times, \eqref{RSD_SIC_5} can be represented as
\begin{equation}
\begin{aligned}
\sum\nolimits_{i=1}^{N_t}{\rm{log}}_2\Big(\Big\lvert\textbf{I}_{N_p}+\frac{N_r}{\sigma^{2}_{k}}\textbf{T}_{i}[k]^{-1}\bm{\bar\Lambda}[k]^2\big(\textbf{B}_{i}[k]\textbf{A}_{i}[k]^H\!+\!\textbf{A}_{i}[k]\textbf{B}_{i}[k]^H\!+\!\textbf{B}_{i}[k]\textbf{B}_{i}[k]^H\big)\Big\rvert\Big),
\end{aligned}
\label{RSR_SIC_expression}
\end{equation}
where $\textbf{A}_{i}[k]=\textbf{C}_{1:i-1}[k]\textbf{S}_{1:i-1}\textbf{F}$, $\textbf{B}_{i}[k]=\textbf{C}_{i}[k]\textbf{S}_{i}\textbf{F}$, and $\textbf{T}_{i}[k]=\textbf{I}_{N_p}+\frac{N_r}{\sigma^{2}_{k}}\bm{\bar\Lambda}[k]^2\textbf{A}_{i}[k]\textbf{A}_{i}[k]^H$. When $i=1$, $\textbf{A}_{1}[k]=\textbf{0}$ and $\textbf{T}_{1}[k]=\textbf{I}_{N_p}$. 
We have now transformed~\eqref{relaxation_SE_1}, which is the $k^{\rm th}$ term of~\eqref{design_problem_1},  to~\eqref{RSR_SIC_expression}. Recall that we aim to design $\textbf{S}$ to maximize~\eqref{design_problem_1}. Hence, designing $\textbf{S}$ to maximize~\eqref{design_problem_1} is transformed as designing $\textbf{S}$ to maximize the summation of~\eqref{RSR_SIC_expression} about $k$, which is given by   
\begin{equation}
\begin{aligned}
\frac{1}{K}\!\sum\nolimits_{i=1}^{N_t}\!\underbrace{\sum\nolimits_{k=1}^{K}\!\!\!{\rm{log}}_2\Big(\Big\lvert\textbf{I}_{N_p}\!\!+\!\!\frac{N_r}{\sigma^{2}_{k}}\textbf{T}_{i}[k]^{-1}\bm{\bar\Lambda}[k]^2\big(\textbf{B}_{i}[k]\textbf{A}_{i}[k]^H\!\!+\!\textbf{A}_{i}[k]\textbf{B}_{i}[k]^H\!\!+\!\textbf{B}_{i}[k]\textbf{B}_{i}[k]^H\big)\Big\rvert\Big).}_{(\star)}
\label{RSR_SIC_expression_1_2}
\end{aligned}
\end{equation}
Note that finding the optimal $\textbf{S}$ to maximize~\eqref{RSR_SIC_expression_1_2} is still difficult due to the summation operation. Fortunately, for each $i$, we observe that $\textbf{T}_i[k]$, $\textbf{A}_i[k]$, and $\textbf{B}_i[k]$ are only related with the first $i^{\rm th}$ rows of $\textbf{S}$ but not related with the remaining rows. According to this property, we propose the following $N_t$-stage method to design $\textbf{S}$ to maximize~\eqref{RSR_SIC_expression_1_2}.
At the first stage, we design $\textbf{S}_1$ to maximize $(\star)$ with $i=1$. At the second stage, with the determined $\textbf{S}_1$ at the first stage, we design $\textbf{S}_2$ to maximize $(\star)$ with $i=2$. Following this trend, at the $N_t^{\rm th}$ stage, the first $N_t-1$ rows of $\textbf{S}$ has been determined and we design $\textbf{S}_{N_t}$ to maximize $(\star)$ with $i=N_t$. After that, all rows of $\textbf{S}$, i.e., the whole $\textbf{S}$, can be determined. 

\begin{table}
	\centering
	\footnotesize
	\begin{tabular}{p{230pt}}
		\hline \textbf{Algorithm 1: RSD algorithm} \\
		\textbf{Input:} $\bm{\bar\Lambda}[k]$, $\textbf{A}_t[k]$, and $\textbf{F}$, $k=1$, $2$, ..., $K$\\
		\quad01:\quad \textbf{for} $i=1:{N_t}$\\
		\quad02:\quad\quad Design $\textbf{S}_i$ by maximizing \eqref{expression_u}\\	
		\quad03:\quad \textbf{end for}\\
		\quad04:\quad Calculate $\textbf{D}[k]$ through \eqref{solution_D_RSD}, for $k=1$, $2$, ..., $K$\\
		\quad05:\quad Normalize each $\textbf{D}[k]$ as $\textbf{D}[k]\leftarrow\frac{\rho}{\sum_{k=1}^{K}\lVert\textbf{S}\textbf{F}\textbf{D}[k]\rVert_F^2}\textbf{D}[k]$\\
		\textbf{Output:} $\textbf{S}$ and $\textbf{D}[k]$, $k=1$, $2$, ..., $K$\\
		\hline
	\end{tabular}
	\vspace{-9.5mm}
\end{table}
Next, we present how to design $\textbf{S}_i$ to maximize $(\star)$ for each stage. Without loss of generality, we use the $i_*^{\rm th}$ stage as an illustration as below.
\begin{subequations}
\begin{align}
&\mathop{\rm max \ }\limits_{\textbf{S}_{i_*}}	\sum\nolimits_{k=1}^{K}\!\!\!{\rm{log}}_2\Big(\Big\lvert\textbf{I}_{N_p}\!\!+\!\!\frac{N_r}{\sigma^{2}_{k}}\textbf{T}_{i_*}[k]^{-1}\bm{\bar\Lambda}[k]^2\big(\textbf{B}_{i_*}[k]\textbf{A}_{i_*}[k]^H\!\!+\!\textbf{A}_{i_*}[k]\textbf{B}_{i_*}[k]^H\!\!+\!\textbf{B}_{i_*}[k]\textbf{B}_{i_*}[k]^H\big)\!\Big\rvert\Big)
\label{problem_RSD_obj}\\ 
&\mathrm{s.t.}\ \textbf{S}_{i_*,l}\in\{0,1\}, \lVert\textbf{S}_{i_*}\rVert_{0}=1,\forall l.
\label{problem_RSD_cons}
\end{align}
\end{subequations}
Due to the structure of $\textbf{F}$ in \eqref{structure_FPS_network}, all the diagonal elements of $\textbf{F}\textbf{F}^{H}$ are $1$. There is only one `1' in $\textbf{S}_{i_*}$ while the other elements are `0's. As a result, regardless how we design $\textbf{S}_{i_*}$, $\textbf{S}_{i_*}\textbf{F}\textbf{F}^{H}\textbf{S}_{i_*}^H$ equals $1$, i.e., $\textbf{B}_{i_*}[k]\textbf{B}_{i_*}[k]^H=\textbf{C}_{i_*}[k]\textbf{S}_{i_*}\textbf{F}\textbf{F}^{H}\textbf{S}_{i_*}^H\textbf{C}_{i_*}[k]^H=\textbf{C}_{i_*}[k]\textbf{C}_{i_*}[k]^H$.
Consequently, by substituting $\textbf{B}_{i_*}[k]=\textbf{C}_{i_*}[k]\textbf{S}_{i_*}\textbf{F}$ and $\textbf{B}_{i_*}[k]\textbf{B}_{i_*}[k]^H=\textbf{C}_{i_*}[k]\textbf{C}_{i_*}[k]^H$ into \eqref{problem_RSD_obj}, \eqref{problem_RSD_obj} can be expressed as
\begin{align}
\sum_{k=1}^{K}\!{\rm{log}}_2\Big(\Big\lvert\textbf{I}_{N_p}\!\!+\!\!\frac{N_r}{\sigma^{2}_{k}}\textbf{T}_{i_*}[k]^{-1}\!\bm{\bar\Lambda}[k]^2\!\big(\textbf{C}_{i_*}\![k]\textbf{S}_{i_*}\textbf{F}\textbf{A}_{i_*}[k]^H\!\!\!+\!\!\textbf{A}_{i_*}[k]\textbf{F}^H\textbf{S}_{i_*}^H\textbf{C}_{i_*}[k]^H\!\!\!+\!\textbf{C}_{i_*}[k]\textbf{C}_{i_*}[k]^H\big)\!\Big\rvert\Big).
\label{expression_u}	
\end{align}
We point out that, since there is only one `1' in $\textbf{S}_{i_*}$, the possible $\textbf{S}_{i_*}$ has $L_tQ$ choices, where $L_tQ$ is the number of FPSs in the DS-FPS architecture. As will be shown in the numerical results in Sec.~\ref{section_simulation}, the number of FPSs is usually limited, e.g., 32. Therefore, it is reasonable and efficient to use the exhaustive search method to find the optimal $\textbf{S}_{i_*}$ to maximize \eqref{expression_u}. The flow and pseudocodes of the RSD algorithm are presented in \textbf{Algorithm 1}.

\section{Low Complexity Row-by-row (RBR) Algorithm}
\label{section_RBR_algorithm}
In the previous section, we have proposed an RSD algorithm to design $\textbf{S}$ and $\textbf{D}[k]$ with the partial CSI. The rows of the RSD algorithm need to be optimized successively, which incurs high complexity. In this section, we propose a low-complexity RBR algorithm, by transforming the design problem into multiple parallel sub-problems and optimizes each row of $\textbf{S}$ in parallel, which effectively reduces the computational complexity.

As analyzed in Sec.~\ref{section_RSD_algorithm}-B, the original design problem can be transformed into problem~\eqref{design_problem_1}. 
By treating $\textbf{S}\textbf{F}\textbf{D}[k]$ as a whole beamforming matrix and considering the low-complexity equal-power allocation, the solution of the optimal unconstrained beamforming matrix to maximize~\eqref{design_problem_1} is $\textbf{P}[k]=\widehat{\textbf{V}}_{N_s}[k]$, where $\widehat{\textbf{V}}_{N_s}[k]$ is the first $N_s$ columns of $\widehat{\textbf{V}}[k]$, and $\widehat{\textbf{V}}[k]$ is derived from the SVD of $\bm{\bar\Lambda}[k]\textbf{A}_t[k]^H$ such that $\bm{\bar\Lambda}[k]\textbf{A}_t[k]^H=\widehat{\textbf{U}}[k]\widehat{\bm{\Sigma}}[k]\widehat{\textbf{V}}[k]^H$.
To reduce the computational complexity, rather than directly solving $\textbf{S}$ and $\textbf{D}[k]$ to maximize~\eqref{design_problem_1}, we propose to design $\textbf{S}$ and $\textbf{D}[k]$ to make $\textbf{S}\textbf{F}\textbf{D}[k]$ close to the optimal unconstrained beamforming matrix $\textbf{P}[k]$, as
\begin{subequations}
	\begin{align}
	&\mathop{\rm min\ }\limits_{\textbf{S},\textbf{D}[k]}\sum\nolimits_{k=1}^{K}\lVert \textbf{P}[k]-\textbf{S}\textbf{F}\textbf{D}[k]\rVert_{F}^2
	\label{problem_SFD_Euclidean_objective}\\ 
	&\mathrm{s.t.}\ \textbf{S}_{i,l}\in\{0,1\}, \lVert\textbf{S}_{i}\rVert_{0}=1,\forall i,l, \sum\nolimits_{k=1}^{K}\lVert\textbf{S}\textbf{F}\textbf{D}[k]\rVert^{2}_{F}=\rho.
	\label{problem_SFD_Euclidean_constraints_2}
	\end{align}
	\label{problem_SFD_Euclidean}%
\end{subequations}
The transformation from design problem~\eqref{design_problem_1} to problem~\eqref{problem_SFD_Euclidean} reduces the complexity, at the cost of an acceptable spectral efficiency loss, as will be shown in the numerical results in Sec.~\ref{section_simulation}. However, it is still uneasy to solve the problem~\eqref{problem_SFD_Euclidean} due to the non-convex binary constraint and the coupling of $\textbf{S}$ and $\textbf{D}[k]$. Hence, we propose the RBR algorithm to alternatively design $\textbf{D}[k]$ and $\textbf{S}$, i.e., alternatively fix one to optimize another one as follows.

\subsection{Design of Digital Beamforming Matrix {\rm \textbf{D}[{\it k}]}}
To begin with, we design $\textbf{D}[k]$ when $\textbf{S}$ is fixed.
A semi-unitary digital beamforming matrix can mitigate the interference among the data streams and enhance the spectral efficiency~\cite{7397861,7579557}.
Inspired by this, we enforce a semi-unitary constraint to the digital beamforming matrix, given by $\textbf{D}[k]^{H}\textbf{D}[k]=\textbf{I}_{N_s}$.
By fixing the switch network matrix $\textbf{S}$ and omitting the transmit power constraint temporarily, the problem~\eqref{problem_SFD_Euclidean} can be reformulated as
\begin{equation}
\begin{aligned}
&{\mathop{\rm  min\ }\limits_{\textbf{D}[k]}}\sum\nolimits_{k=1}^{K}\lVert \textbf{P}[k]-\textbf{S}\textbf{F}\textbf{D}[k]\rVert_{F}^2\\ 
&\mathrm{s.t.}\ \textbf{D}[k]^{H}\textbf{D}[k]=\textbf{I}_{N_s}.
\end{aligned}
\label{problem_SFD_Euclidean_D}
\end{equation}
The solution to~\eqref{problem_SFD_Euclidean_D}, which is the orthogonal procrustes problem, is given as~\cite{7397861,7579557} 
\begin{equation}
\textbf{D}[k]=\ddot{\textbf{V}}_{N_s}[k]\ddot{\textbf{U}}[k]^{H},
\label{solution_D_RBR}
\end{equation}
where $L_t\times L_t$- and $N_s\times N_s$-dimensional $\ddot{\textbf{V}}[k]$ and $\ddot{\textbf{U}}[k]$ are obtained from the SVD of $\textbf{P}[k]^{H}\textbf{S}\textbf{F}$, yielding that $\textbf{P}[k]^{H}\textbf{S}\textbf{F}=\ddot{\textbf{U}}[k]\ddot{\bm \Sigma}[k]\ddot{\textbf{V}}[k]^{H}$, and $\ddot{\textbf{V}}_{N_s}[k]$ is the first $N_s$ columns of $\ddot{\textbf{V}}[k]$. 
\subsection{Design of Switch Network Matrix {\rm \textbf{S}}}
Then, we design $\textbf{S}$ to solve the problem \eqref{problem_SFD_Euclidean}, with fixed $\textbf{D}$.
By omitting the transmit power constraint temporarily, solving $\textbf{S}$ to minimize \eqref{problem_SFD_Euclidean_objective} is rearranged as
\begin{subequations}
	\begin{align}
	&\mathop{\rm min\ }\limits_{\textbf{S}}\sum\nolimits_{k=1}^{K}\lVert \textbf{P}[k]-\textbf{S}\textbf{F}\textbf{D}[k]\rVert_{F}^2
	\label{subproblem_SFD_Euclidean_objective}\\& 
	\mathrm{s.t.}\ \textbf{S}_{i,l}\in\{0,1\}, \
	\lVert\textbf{S}_{i}\rVert_{0}=1, \forall i,l,
	\label{subproblem_SFD_Euclidean_constraint_1}
	\end{align}
	\label{subproblem_SFD_Euclidean}%
\end{subequations}
where~\eqref{subproblem_SFD_Euclidean} is an integer programming problem associated with a matrix variable, which is inefficient to solve. To make the problem more tractable, we rewrite the $k^{\rm th}$ term of \eqref{subproblem_SFD_Euclidean_objective} as
\begin{subequations}
	\begin{align}
	&\quad \ \left\lVert \textbf{P}[k]-\textbf{S}\textbf{F}\textbf{D}[k]\right\rVert_{F}^2\label{trace_expression_1}\\
	&={\rm Tr}\left((\textbf{P}[k]-\textbf{S}\textbf{F}\textbf{D}[k])(\textbf{P}[k]-\textbf{S}\textbf{F}\textbf{D}[k])^H\right)\label{trace_expression_2}\\
	&={\rm Tr}\left(\textbf{P}[k]\textbf{P}[k]^H\right)+{\rm Tr}\left(\textbf{S}\textbf{F}\textbf{D}[k]\textbf{D}[k]^H\textbf{F}^H\textbf{S}^H\right)-2{\rm Tr}\left({\rm Re}(\textbf{S}\textbf{F}\textbf{D}[k]\textbf{P}[k]^H)\right)\label{trace_expression_3}\\
	&={\rm Tr}\left(\textbf{P}[k]\textbf{P}[k]^H\right)+{\rm Tr}\bigg(\textbf{S}\textbf{F}\textbf{K}[k]\bigg[\!\!\begin{array}{cc}
	\textbf{I}_{N_s}&\\
	&\bm{0}\\
	\end{array}\!\!\bigg]\textbf{K}[k]^H\textbf{F}^H\textbf{S}^H\bigg)-2{\rm Tr}\left({\rm Re}(\textbf{S}\textbf{F}\textbf{D}[k]\textbf{P}[k]^H)\right)\label{trace_expression_4}\\
	&={\rm Tr}\left(\textbf{P}[k]\textbf{P}[k]^H\right)+{\rm Tr}\bigg(\bigg[\!\!\begin{array}{cc}
	\textbf{I}_{N_s}&\\
	&\bm{0}\\
	\end{array}\!\!\bigg]\textbf{K}[k]^H\textbf{F}^H\textbf{S}^H\textbf{S}\textbf{F}\textbf{K}[k]\bigg)-2{\rm Tr}\left({\rm Re}(\textbf{S}\textbf{F}\textbf{D}[k]\textbf{P}[k]^H)\right)\label{trace_expression_5}\\
	&\leq{\rm Tr}\left(\textbf{P}[k]\textbf{P}[k]^H\right)+{\rm Tr}(\textbf{K}[k]\textbf{K}[k]^H\textbf{F}^H\textbf{S}^H\textbf{S}\textbf{F})-2{\rm Tr}\left({\rm Re}(\textbf{S}\textbf{F}\textbf{D}[k]\textbf{P}[k]^H)\right)\label{trace_expression_6}\\
	&={\rm Tr}\left(\textbf{P}[k]\textbf{P}[k]^H\right)+{\rm Tr}(\textbf{F}^H\textbf{S}^H\textbf{S}\textbf{F})-2{\rm Tr}\left({\rm Re}(\textbf{S}\textbf{F}\textbf{D}[k]\textbf{P}[k]^H)\right),
	\label{trace_expression_7}
	\end{align}
\end{subequations}
where $\textbf{K}[k]\bigg[\!\!\begin{array}{cc}
\textbf{I}_{N_s}&\\
&\bm{0}\\
\end{array}\!\!\bigg]\textbf{K}[k]^H$ in \eqref{trace_expression_4} is the SVD of $\textbf{D}[k]\textbf{D}[k]^H$ since we have $\textbf{D}[k]^H\textbf{D}[k]=\textbf{I}_{N_s}$. \eqref{trace_expression_5} comes from the property of matrix trace. The inequality \eqref{trace_expression_6} follows that the diagonal elements of the Hermitian matrix $\textbf{K}[k]^H\textbf{F}^H\textbf{S}^H\textbf{S}\textbf{F}\textbf{K}[k]$ are no smaller than zero and the equality holds when $\textbf{D}[k]$ is a square matrix, i.e., $L_t=N_s$. Therefore, the $k^{\rm th}$ term of~\eqref{subproblem_SFD_Euclidean_objective} can be relaxed as \eqref{trace_expression_7}, where ${\rm Tr}\left(\textbf{P}[k]\textbf{P}[k]^H\right)$ is known and fixed. According to the structure of $\textbf{F}$ in \eqref{structure_FPS_network} and the constraint $\lVert\textbf{S}_i\rVert_{0}=1$, regardless how we design $\textbf{S}$, ${\rm Tr}(\textbf{F}^H\textbf{S}^H\textbf{S}\textbf{F})=\lVert\textbf{S}\textbf{F}\rVert_F^2$ is a constant $N_t$. Hence, to minimize~\eqref{trace_expression_7} is equivalent to maximize ${\rm Tr}\left({\rm Re}(\textbf{S}\textbf{F}\textbf{D}[k]\textbf{P}[k]^H)\right)$. Note that~\eqref{trace_expression_7} is a relaxed form of the $k^{\rm th}$ term of~\eqref{subproblem_SFD_Euclidean_objective}. Therefore, minimizing~\eqref{subproblem_SFD_Euclidean_objective} can be relaxed as maximizing the summation of ${\rm Tr}\left({\rm Re}(\textbf{S}\textbf{F}\textbf{D}[k]\textbf{P}[k]^H)\right)$ about $k$, given by $\sum\nolimits_{k=1}^{K}{\rm Tr}\left({\rm Re}(\textbf{S}\textbf{F}\textbf{D}[k]\textbf{P}[k]^H)\right)$. According to the property of matrix trace, $\sum\nolimits_{k=1}^{K}{\rm Tr}\left({\rm Re}(\textbf{S}\textbf{F}\textbf{D}[k]\textbf{P}[k]^H)\right)$ is equivalent to 
\begin{equation}
\sum\nolimits_{i=1}^{N_t}\sum\nolimits_{k=1}^{K}{\rm Re}(\textbf{S}_i\textbf{F}\textbf{D}[k]\textbf{P}_i[k]^H),
\label{third_term_1}
\end{equation}
where $\textbf{S}_i$ and $\textbf{P}_i[k]$ are the $i^{\rm th}$ row of $\textbf{S}$ and $\textbf{P}[k]$, respectively. According to~\eqref{third_term_1}, we have decomposed each row of $\textbf{S}$ as $N_t$ uncorrelated parts. As a result, designing $\textbf{S}$ to maximize~\eqref{third_term_1} is equivalent to separately designing $\textbf{S}_i$ to maximize $\sum\nolimits_{k=1}^{K}{\rm Re}(\textbf{S}_i\textbf{F}\textbf{D}[k]\textbf{P}_i[k]^H)$, for $i=1$, $2$, ..., $N_t$, which can be stated as
\begin{subequations}
	\begin{align}
	&\mathop{\rm max\ }\limits_{\textbf{S}_i}\textbf{S}_i\sum\nolimits_{k=1}^{K}{\rm Re}(\textbf{F}\textbf{D}[k]\textbf{P}_i[k]^H)
	\label{subproblem_SFD_2_obj}
	\\& 
	\mathrm{s.t.}\ \textbf{S}_{i,l}\in\{0,1\},\ \lVert\textbf{S}_{i}\rVert_{0}=1, \forall l.
	\end{align}
	\label{subproblem_SFD_2}%
\end{subequations}
Following the binary property of the row vector $\textbf{S}_i$, i.e., $\textbf{S}_{i,l}\in\{0,1\}, \forall l$ and $ \lVert\textbf{S}_{i}\rVert_{0}=1$, maximizing \eqref{subproblem_SFD_2_obj} is equivalent to finding the position of the maximal element of the column vector $\sum\nolimits_{k=1}^{K}{\rm Re}(\textbf{F}\textbf{D}[k]\textbf{P}_i[k]^H)$, which is a simple ranking problem and can be solved efficiently based on the sorting algorithm in solvers. Consequently, by denoting $p_{\rm max}$ as the position of the maximal element, the optimal solution of $\textbf{S}_i$ to the problem \eqref{subproblem_SFD_2} is 
\begin{equation}
\textbf{S}_i=[\ \underbrace{0,...,0}_{p_{\rm max}-1},\ 1,\underbrace{0,...,0}_{L_tQ-p_{\rm max}}].
\label{solution_S_RBR}
\end{equation}
Then, by solving problem \eqref{subproblem_SFD_2} with $i=1$, ..., $N_t$ in parallel, the solution of $\textbf{S}$ to problem \eqref{subproblem_SFD_Euclidean} is obtained. Based on the aforementioned procedures, in the proposed RBR algorithm, we can alternatively solve $\textbf{D}[k]$ and $\textbf{S}$ via \eqref{solution_D_RBR} and \eqref{solution_S_RBR} until convergence. After that, we enforce the transmit power constraint to $\textbf{D}[k]$ such 
that $\textbf{D}[k]\leftarrow\frac{\rho}{\sum_{k=1}^{K}\lVert\textbf{S}\textbf{F}\textbf{D}[k]\rVert_F^2}\textbf{D}[k]$.
The pseudocodes of the RBR algorithm are described in \textbf{Algorithm 2}. 

\begin{table}
	\centering
	\footnotesize
	\begin{tabular}{p{260pt}}
		\hline \textbf{Algorithm 2: RBR algorithm} \\
		\textbf{Input:} $\bm{\bar\Lambda}[k]$, $\textbf{A}_t[k]$, and $\textbf{F}$, $k=1$, $2$, ..., $K$\\
		\quad01:\quad Calculate each $\textbf{P}[k]=\widehat{\textbf{V}}_{N_s}[k]$ and initialize $\textbf{S}$ randomly\\							
		\quad02:\quad \textbf{Repeat}\\
		\quad03:\quad Solve $\textbf{D}[k]$ via~\eqref{solution_D_RBR}, for $k=1$, $2$, ..., $K$\\		
		\quad04:\quad\ \ \textbf{For} $i=1:{N_t}$\\
		\quad05:\quad\quad\ \ Solve $\textbf{S}_i$ via~\eqref{solution_S_RBR}  \\			
		\quad06:\quad\ \ \textbf{end for}\\	
		\quad07:\quad \textbf{Until convergence} \\
		\quad08:\quad Normalize each $\textbf{D}[k]$ as  $\textbf{D}[k]\leftarrow\frac{\rho}{\sum_{k=1}^{K}\lVert\textbf{S}\textbf{F}\textbf{D}[k]\rVert_F^2}\textbf{D}[k]$\\
		\textbf{Output:} $\textbf{S}$ and $\textbf{D}[k]$, $k=1$, $2$, ..., $K$\\
		\hline
	\end{tabular}
	\vspace{-9.5mm}
\end{table}

\section{Simulation Results and Analysis}
\label{section_simulation}   
In this section, we evaluate the performance of the proposed DS-FPS architecture as well as the RSD and RBR algorithms. The simulation setup is given in Sec.~\ref{section_simulation}-A. We first evaluate the spectral efficiency and energy efficiency of the DS-FPS architecture with the RSD and RBR algorithms in Sec.~\ref{section_simulation}-B. Then, we analyze the impact of CSI on the spectral efficiency for the DS-FPS architecture in Sec.~\ref{section_simulation}-C. Furthermore, we analyze the computational complexity and convergence of the RSD and RBR algorithms in Sec.~\ref{section_simulation}-D.

\subsection{Simulation Setup}
\subsubsection{\textbf{Generation of THz UM-MIMO Channel}}
\begin{figure}
	\setlength{\belowcaptionskip}{0pt}
	\centering
	\includegraphics[scale=0.3]{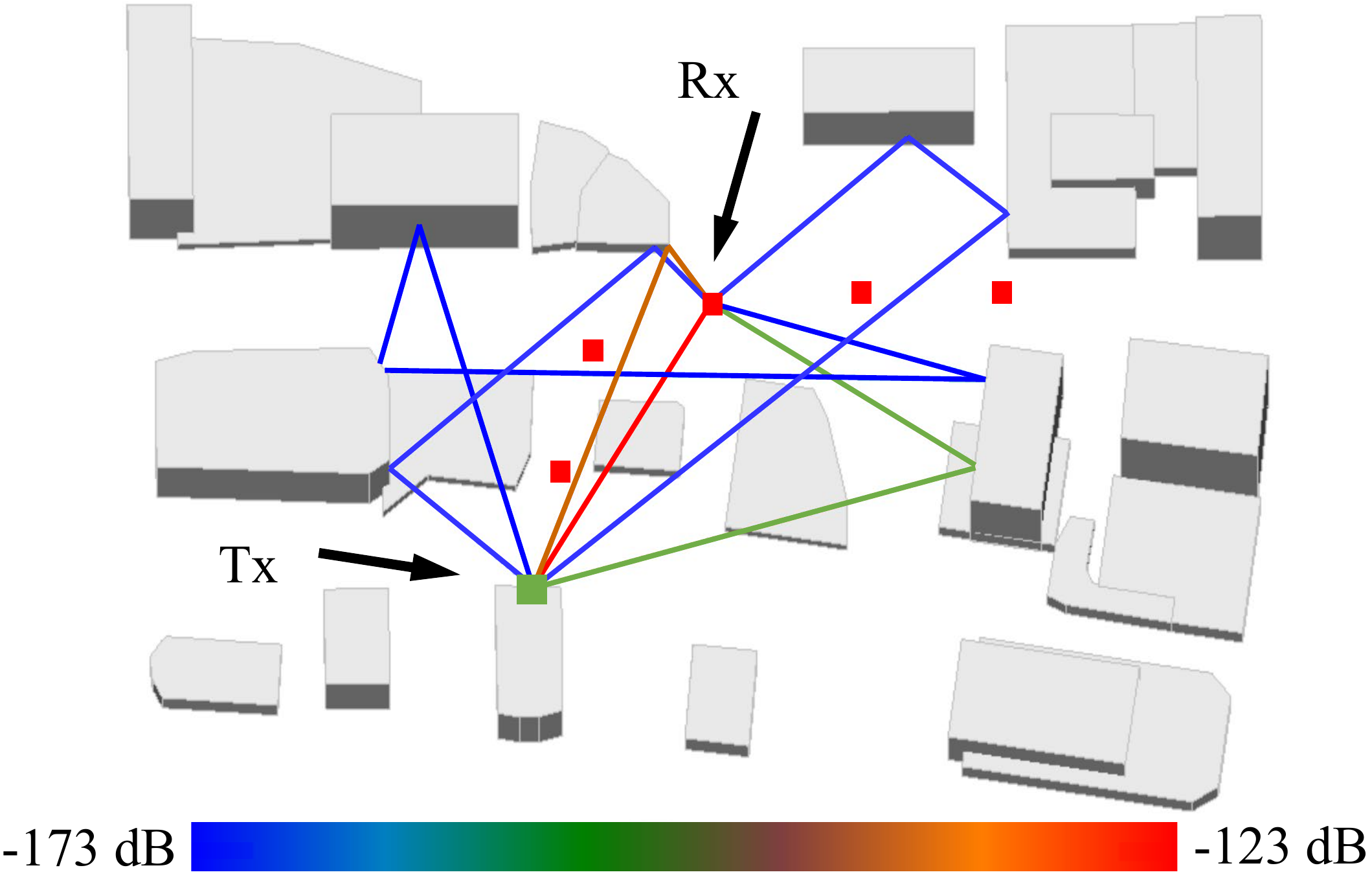} 
	\captionsetup{font={footnotesize}}
	\caption{The environment of THz channels.
	}  
	\label{fig_channel_setup}
	\vspace{-9.5mm}
\end{figure}

The operating frequency is $0.3$ THz, with $5$~GHz bandwidth. The number of sub-carrier is 10 and the noise power is $-87$ dBm for each sub-carrier.
We consider a typical outdoor street scenario as shown in Fig.~\ref{fig_channel_setup}. The transmitter (TX) is arranged at the roof of one building, whose height is $30$m. The height of the receiver (RX) is 1.5m. There are five positions of RX in red points, for which the LoS distances $D$ equal $40$m, $70$m, $100$m, $130$m, and $160$m. The number of antennas at TX and RX is equal, which is set as $128$, $256$, $512$, and $1024$, respectively.
The ray-tracing tool Wireless Insite is adopted, which can characterize the multipath components with high accuracy, to calculate the parameters of each propagation path from TX to RX, e.g., the DoD and DoA~\cite{8304810}. Then, we compute the path gains according to our previous THz channel work~\cite{6998944}, by considering the THz-specific propagation characteristics. For illustration, the lines with different colors in~Fig.~\ref{fig_channel_setup} denote the propagation paths and path gains with $100$m LoS separation at $0.3$ THz. The paths whose path gain is weaker than the LoS path over 50 dB are omitted since their contributions to the channel are negligible. Substituting the path gains, DoA, and DoD into~\eqref{channel_model_planar}, we can obtain the THz UM-MIMO channel.

\subsubsection{\textbf{Competitors of Proposed DS-FPS Architecture}} We compare our scheme, i.e., the proposed DS-FPS architecture with RSD and RBR algorithms, with the following schemes: i) the FC architecture with the hybrid beamforming algorithm in~\cite{7913599}, ii) the DHB architecture with the algorithm in~\cite{9110865}, iii) the FPS group-connected (FPS-GC) architecture with the FPS-AltMin algorithm in~\cite{8310586}, iv) the AoSA architecture with SIC algorithm in~\cite{7445130}, 
v) the subconnected phase shifter network with fully connected switch networks (SPSF) architecture with the algorithm in~\cite{8295113}, and iv) the subconnected with reduced number of phase shifters and phase shifter selection (SRPS) architecture with the algorithm in~\cite{8295113,8382230}. For the FPS-GC architecture, the number of groups is set as 2 to achieve high energy efficiency. 
The key idea of the SPSF and SRPS architectures is using switches to dynamically turn off partial antennas with relatively small contribution to the spectral efficiency. The number of phase shifters is $\frac{N_t}{\beta}$, where $N_t$ is the number of antennas and $1-\frac{1}{\beta}$ is the ratio of the antennas which need to be turned off. Specifically, in SPSF architecture, each phase shifter connects to $\frac{N_t}{L_t}$ adjacent antennas through $\frac{N_t}{L_t}$ switches, where $L_t$ is the number of RF chains. Among these $\frac{N_t}{L_t}$ antennas, $\frac{N_t}{L_t}(1-\frac{1}{\beta})$ antennas are turned off. In SRPS architecture, each phase shifter connects to $\beta$ adjacent antennas through 1-to-$\beta$ switch and then remains one antenna with largest contribution to the spectral efficiency. In this work, we set $\beta=2$ for both SPSF and SRPS architectures.
The major difference between our DS-FPS architecture and the SPSF and SRPS architectures is that the SPSF and SRPS use adjustable phase shifters while the DS-FPS uses low-cost FPSs. Moreover, the switches in the DS-FPS architecture are used to allow each antenna selecting one proper FPS such that all antennas are active, while in the SPSF and SRPS architectures partial antennas are turned off. With different hardware architectures, the algorithms in this work are also different from those in~\cite{8295113,8382230}.
For all architectures, the number of RF chains at TX and RX is $L_t=4$. The number of data streams is $N_s=4$.  
In our scheme, we adopt the DS-FPS architecture at both transmitter and receiver with the same system parameters. For fair comparison, the counterpart architectures are also used at two sides.

\subsubsection{\textbf{Energy Efficiency Model}}
\begin{table}
	\centering
	\footnotesize
	\captionsetup{font={footnotesize}}
	\caption{Power consumption at transmitter of different architectures.}
	\begin{tabular}{cc} 
		\hline
		Architecture&Power consumption\\
		\hline 
		Proposed DS-FPS&${\rm P_{common}}+{\rm P_{SW}}N_t+{\rm P_{FPS}}{\rm N_{FPS}^{a}}$\\
		FC&${\rm P_{common}}+{\rm P_{PS}}N_{t}L_t$\\
		AoSA&${\rm P_{common}}+{\rm P_{PS}}N_{t}$\\	
		DHB&${\rm P_{common}}+{\rm P_{SW}}N_{t}+{\rm P_{PS}}N_{t}$\\
		FPS-GC&${\rm P_{common}}+{\rm P_{SW}}{\rm N_{SW}}+{\rm P_{FPS}}{\rm N_{FPS}}$\\
		SPSF \& SRPS&${\rm \widehat{P}_{common}}+{\rm P_{PS}}N_{t}/\beta+{\rm P_{SW}}N_{t}/\beta$\\
		\hline	
	\end{tabular}
	\label{table_power_existing_architectures}
	\vspace{-7.5mm}
\end{table}
The energy efficiency $EE$ is defined as the ratio between the spectral efficiency and the power consumption at transmitter $P_{TX}$ and receiver $P_{RX}$, i.e., $EE=\frac{SE}{P_{TX}+P_{RX}}$.
We adopt the power consumption model of the hybrid beamforming studies~\cite{9110865,9374093}, according to which the power consumption of the DS-FPS architecture at transmitter can be expressed as
\begin{equation}
P_{TX}={\rm P_{common}}+\underbrace{{\rm P_{SW}}N_t+{\rm P_{FPS}}{\rm N_{FPS}^a}}_{\rm P_{analog}},
\label{eq_power_consumption_DS_FPS}
\end{equation}
where ${\rm P_{common}}={\rm P_{BB}}+{\rm P_{DAC}}L_t+{\rm P_{RF}}L_t+{\rm P_{PA}}N_t+\rho$. ${\rm P_{BB}}$, ${\rm P_{DAC}}$, ${\rm P_{RF}}$, ${\rm P_{PA}}$, ${\rm P_{SW}}$, ${\rm P_{FPS}}$ denote the power consumption of the baseband, DAC, RF chain, power amplifier, switch, and FPS, respectively. The corresponding multipliers denote the quantity of these devices used in the architecture. $\rho$ is the transmit power at transmitter. ${\rm P_{common}}$ is usually the same for different hybrid beamforming architectures.
${\rm P_{analog}}$ is the analog beamforming part of power consumption, which is different for various hybrid beamforming architectures according to the used hardware components to realize the analog beamforming. For the DS-FPS architecture, ${\rm P_{analog}}$ is composed by the power consumed by switches and FPSs as expressed in~\eqref{eq_power_consumption_DS_FPS}.

In the DS-FPS architecture, the total number of FPSs is $L_tQ$. Each antenna connects to all FPSs through $L_tQ$ switches. As analyzed in Sec.~\ref{section_channel_system_model}-B, each antenna selects one FPS with proper phase from all FPSs to generate the beamforming weight. For one antenna, only one of the $L_tQ$ switches is closed and the others are disconnected which do not consume power. Hence, the number of switches which consume power equals the number of antennas $N_t$.
In this work, $N_t$, $L_t$, and $\rho$ are not design variables such that ${\rm P_{common}}+{\rm P_{SW}}N_t$ can be referred to as the static terms which are note related to the hybrid beamforming design. We denote the FPSs which are selected by at least one switch as the active FPSs and the others as the non-active FPSs which do not consume power. The number of active FPSs ${\rm N_{FPS}^a}$ equals the number of non-zero column of $\textbf{S}$. Hence, ${\rm P_{FPS}}{\rm N_{FPS}^a}$ is related to the design variable $\textbf{S}$ and is the dynamic term.

For FC, AoSA, DHB, and FPS-GC architectures, the common part is the same with~\eqref{eq_power_consumption_DS_FPS}, while the analog beamforming part needs to be changed by accounting their own devices. In FPS-GC architecture, ${\rm N_{SW}}$ and ${\rm N_{FPS}}$ should be determined by the FPS-AltMin algorithm~\cite{8310586}. For SPSF and SRPS architectures, only $N_t/\beta$ antennas are active such that ${\rm \widehat{P}_{common}}={\rm P_{BB}}+{\rm P_{DAC}}L_t+{\rm P_{RF}}L_t+{\rm P_{PA}}N_t/\beta+\rho$. For the analog beamforming part, the number of phase shifters is $N_t/\beta$. Moreover, each phase shifter only has one closed switch which consumes power such that the number of closed switches is $N_t/\beta$. Hence, the consumed power of analog beamforming part is ${\rm P_{PS}}N_{t}/\beta+{\rm P_{SW}}N_{t}/\beta$. We summarize the power consumption at transmitter, i.e., $P_{TX}$, of different architectures in~TABLE~\ref{table_power_existing_architectures}. 
For each architecture, the expression of $P_{RX}$ is similar with $P_{TX}$, by substituting the DAC and power amplifier with ADC and low noise power and then deleting $\rho$.

We use the typical power consumption reported by the existing THz studies, mainly around 0.3 THz, in the unit of mW as follows. The power consumption of DAC, ADC, power amplifier, low noise amplifier, switch, RF chain, and baseband is ${\rm P_{DAC}}=110$~\cite{7019002}, ${\rm P_{ADC}}=158.6$~\cite{8951264}, ${\rm P_{PA}}=49$~\cite{5518016}, ${\rm P_{LNA}}=53$~\cite{5619646}, ${\rm P_{SW}}=9$~\cite{9223920}, ${\rm P_{RF}}=43$~\cite{8733134}, and ${\rm P_{BB}}=200$~\cite{7436794}, respectively. It has been reported in~\cite{8058787} that ${\rm P_{PS}}=52$, $39$, and $26$ for 3, 2, and 1-bit phase shifters. The FC~\cite{7913599}, AoSA~\cite{7445130}, SPSF~\cite{8295113}, and SRPS~\cite{8382230} architectures use the infinite-resolution phase shifter, while the DHB architecture~\cite{9110865} uses the 2-bit or 1-bit resolution phase shifters. Since the infinite-resolution phase shifter is ideal, we use the power consumption of 3-bit phase shifter to represent its power consumption. Hence, for FC, AoSA, SPSF, and SRPS architectures, ${\rm P_{PS}}=52$, while for DHB architecture, ${\rm P_{PS}}=39$ for 2-bit phase shifter and ${\rm P_{PS}}=26$ for 1-bit phase shifter. 
The FPS is usually realized by a passive delay element which provides a fixed phase adjustment with no power consumption and an amplifier with small gain to keep the output amplitude and power of the FPSs with different phases at the same level~\cite{9426936}.
Since the passive delay element with different phases in~\cite{9426936} has about 3.3 dB output power difference, we consider an amplifier with about 4.8 dB gain which consumes 16.8 mW~\cite{4633188} to keep the equal output amplitude and power of FPSs. Hence, the power consumption of FPS is ${\rm P_{FPS}}=16.8$.

In this work, we consider the use of full-bit DACs/ADCs. It has been studied in~\cite{9205899} that we can use the power consumption of DAC/ADC ($>$8 bit) to represent the power consumption of full-bit DAC/ADC, for which we use the power consumption of 9-bit DAC in~\cite{7019002}, i.e., ${\rm P_{DAC}}=110$ mW, and 12-bit ADC in~\cite{8951264}, i.e., ${\rm P_{ADC}}=158.6$ mW.
The use of low-bit DACs/ADCs may be a potential research direction. The joint optimization of the hybrid beamforming matrices and the bit of the DACs/ADCs has been analyzed in~\cite{9205899} and a channel estimation algorithm for hybrid beamforming with low-bit ADCs has been proposed in~\cite{8553303}.

\subsection{Spectral Efficiency and Energy Efficiency of the Proposed DS-FPS Architecture}
In the following simulations, the DS-FPS architecture with the RSD and RBR algorithms only knows partial CSI. For the other architectures and algorithms, full CSI is assumed to be known.

\begin{figure}
	\setlength{\belowcaptionskip}{0pt}
	\centering
	\captionsetup{font={footnotesize}}
	\subfigure[Spectral efficiency versus transmit power $\rho$.]{
		\includegraphics[scale=0.28]{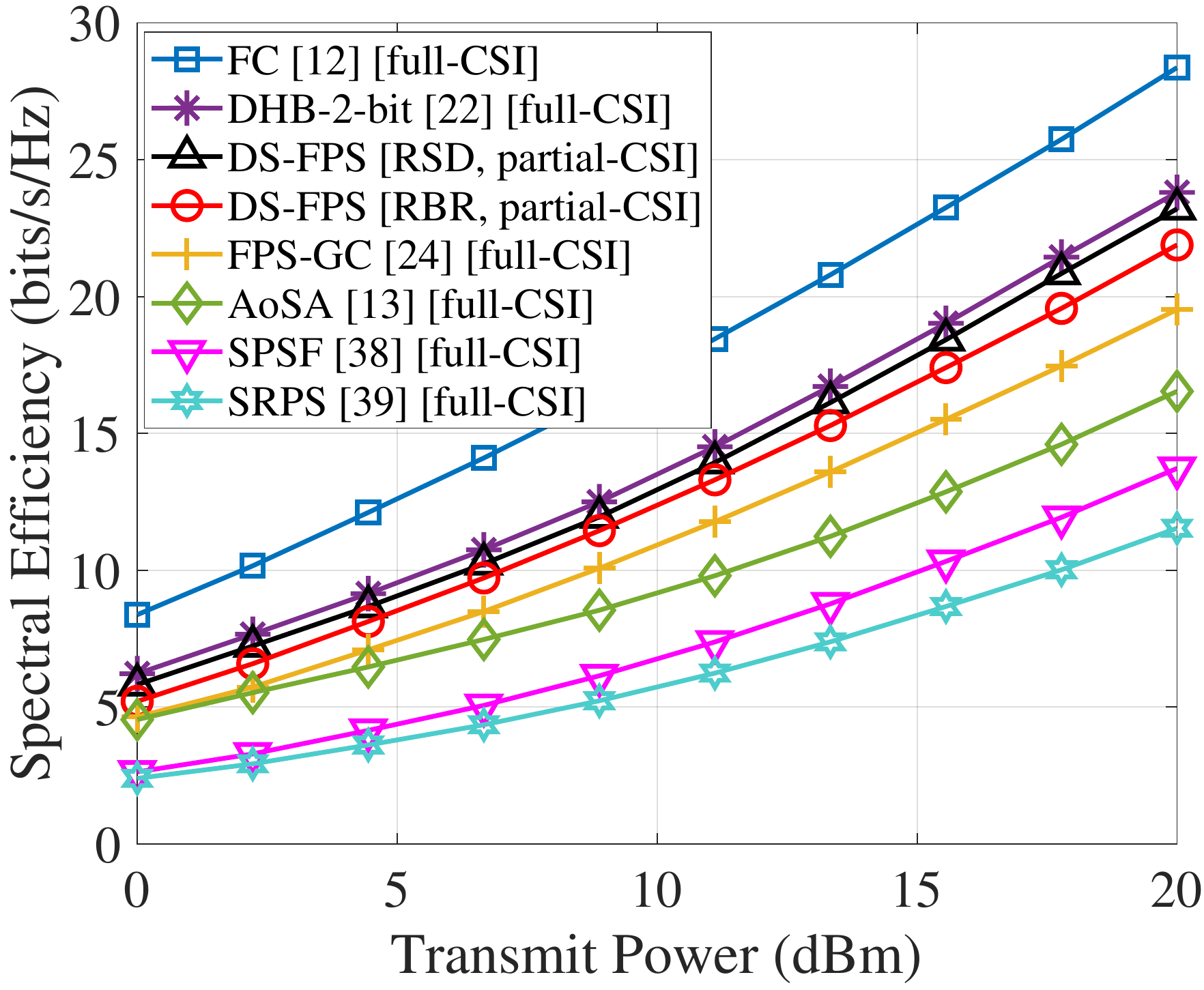}}
	\subfigure[Energy efficiency versus spectral efficiency, $\rho=20$~dBm.]{
		\includegraphics[scale=0.28]{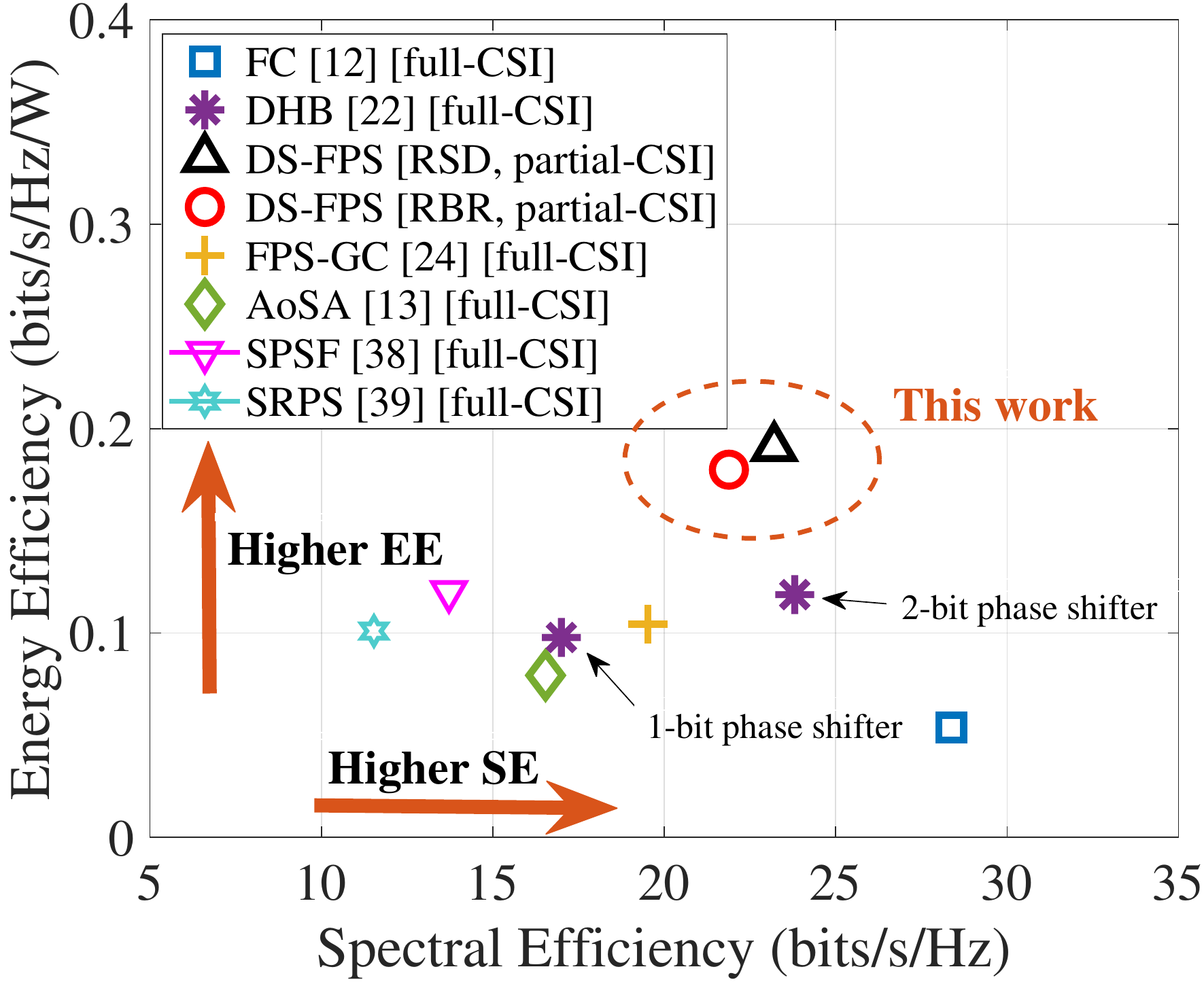}}
	\subfigure[Energy efficiency versus spectral efficiency with another group of power consumption values, $\rho=20$~dBm.]{
		\includegraphics[scale=0.28]{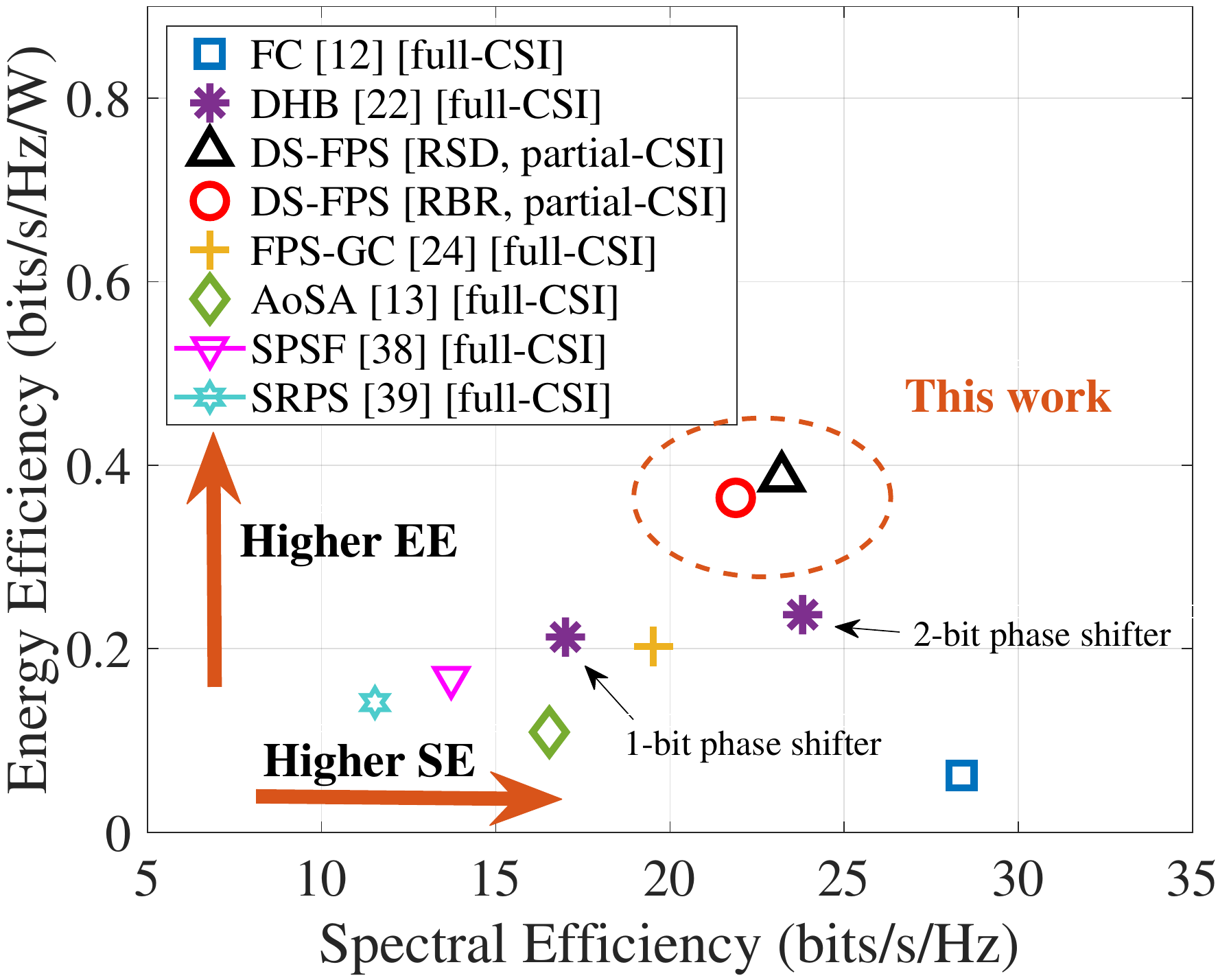}}
	\caption{The spectral efficiency and energy efficiency of the DS-FPS architecture. Communication distance is $D=40$m, $N_t=N_r=1024$. The number of FPSs in DS-FPS and FPS-GC is $32$.}
	\label{fig_SE_EE_architecture}
	\vspace{-9.5mm}
\end{figure}

Fig.~\ref{fig_SE_EE_architecture}(a) and Fig.~\ref{fig_SE_EE_architecture}(b) evaluate the spectral efficiency and energy efficiency of the proposed DS-FPS architecture. As shown in Fig.~\ref{fig_SE_EE_architecture}(a), for the DS-FPS architecture, the RSD algorithm yields higher spectral efficiency than the low-complexity RBR algorithm, e.g., $1.4$ bits/s/Hz when $\rho=20$ dBm.
Furthermore, both the spectral efficiencies of the DS-FPS architecture with either RSD or RBR algorithm are higher than the FPS-GC and the AoSA architectures, e.g., $2$ bits/s/Hz and $6$ bits/s/Hz when $\rho=20$~dBm. The SPSF and SRPS architectures aim to reduce the power consumption by deactivating partial antennas, which decreases the spectral efficiency at the same time. As a result, the spectral efficiency of the DS-FPS architecture is about 9~bits/s/Hz and 11~bits/s/Hz higher than the SPSF and SRPS architectures when $\rho=20$~dBm. The SPSF has higher spectral efficiency than the SRPS due to the more flexible switch connections.
Furthermore, the spectral efficiencies of the DS-FPS architecture with the RSD and RBR algorithms are similar to the DHB with 2-bit phase shifter and lower than the FC.  
Therefore, the proposed DS-FPS architecture with the RSD and RBR algorithms can achieve good spectral efficiency.

Fig.~\ref{fig_SE_EE_architecture}(b) evaluates the energy efficiency as well as the spectral efficiency concurrently. The energy efficiencies of the DS-FPS architecture with the RSD and RBR algorithms are significantly higher than the other architectures, due to the low power consumption of the FPS and switch. The power consumed by switches is only a small part of the overall power consumption of DS-FPS architecture, e.g., 15\% in Fig.~\ref{fig_SE_EE_architecture}(b). Although the spectral efficiency of the DS-FPS architecture is lower than the FC architecture, the energy efficiency of the DS-FPS architecture is 
remarkably higher than the FC, i.e., over $3$ times. The DHB architecture aims to use low-cost 2-bit and 1-bit phase shifter to reduce the power consumption and achieve better trade-off between energy efficiency and spectral efficiency. However, the power consumption of 2-bit and 1-bit phase shifter is still higher than the FPS in this work. Moreover, the quantity of phase shifters in DHB architecture equals $N_t$, i.e., $1024$, which is much larger than the number of FPSs in DS-FPS architecture, i.e., $32$. Hence, the power consumption of the DS-FPS architecture is much lower than the DHB architecture. Compared to the DHB architecture with 2-bit phase shifter, the proposed DS-FPS architecture can achieve similar spectral efficiency, while with $60$\% energy efficiency enhancement. Both the spectral efficiency and energy efficiency of the DS-FPS architecture are higher than the DHB architecture with 1-bit phase shifter. The spectral efficiency and energy efficiency of the proposed DS-FPS architecture are noticeably higher than the FPS-GC and AoSA architectures. With $\beta=2$, the SPSF and SRPS architectures deactivate half antennas, which reduces the power consumption substantially. However, with half antennas, the spectral efficiency is also low, which makes that the energy efficiency is lower than the DS-FPS architecture.
Therefore, by using the RSD and RBR algorithms, the proposed DS-FPS architecture can achieve substantially improved energy efficiency and good spectral efficiency at the same time, compared to the existing competitors.

In Fig.~\ref{fig_SE_EE_architecture}(c), we evaluate the impact of different choice of power consumption value (unit:~mW) on energy efficiency, compared to the value provided in Sec.~\ref{section_simulation}-A-3). We set ${\rm P_{DAC}}=110$, ${\rm P_{ADC}}=200$, ${\rm P_{PA}}=16$, ${\rm P_{LNA}}=30$, ${\rm P_{RF}}=43$, and ${\rm P_{BB}}=243$, as the literature~\cite{8733134}. We set ${\rm P_{SW}}=5$, ${\rm P_{PS}}=50$, $20$, $10$ for infinite-resolution, 2-bit, and 1-bit phase shifter, respectively, as the literature~\cite{9110865}. Moreover, we set ${\rm P_{FPS}}=6.8$~\cite{9360389}. Although the detailed values of the energy efficiency change substantially, the energy efficiency of the DS-FPS architecture is much higher than the others, which is similar to Fig.~\ref{fig_SE_EE_architecture}(b).

\begin{figure}
	\centering
	\begin{tabular}{cc}
		\begin{minipage}[t]{0.32\linewidth}
			\includegraphics[width = 1\linewidth]{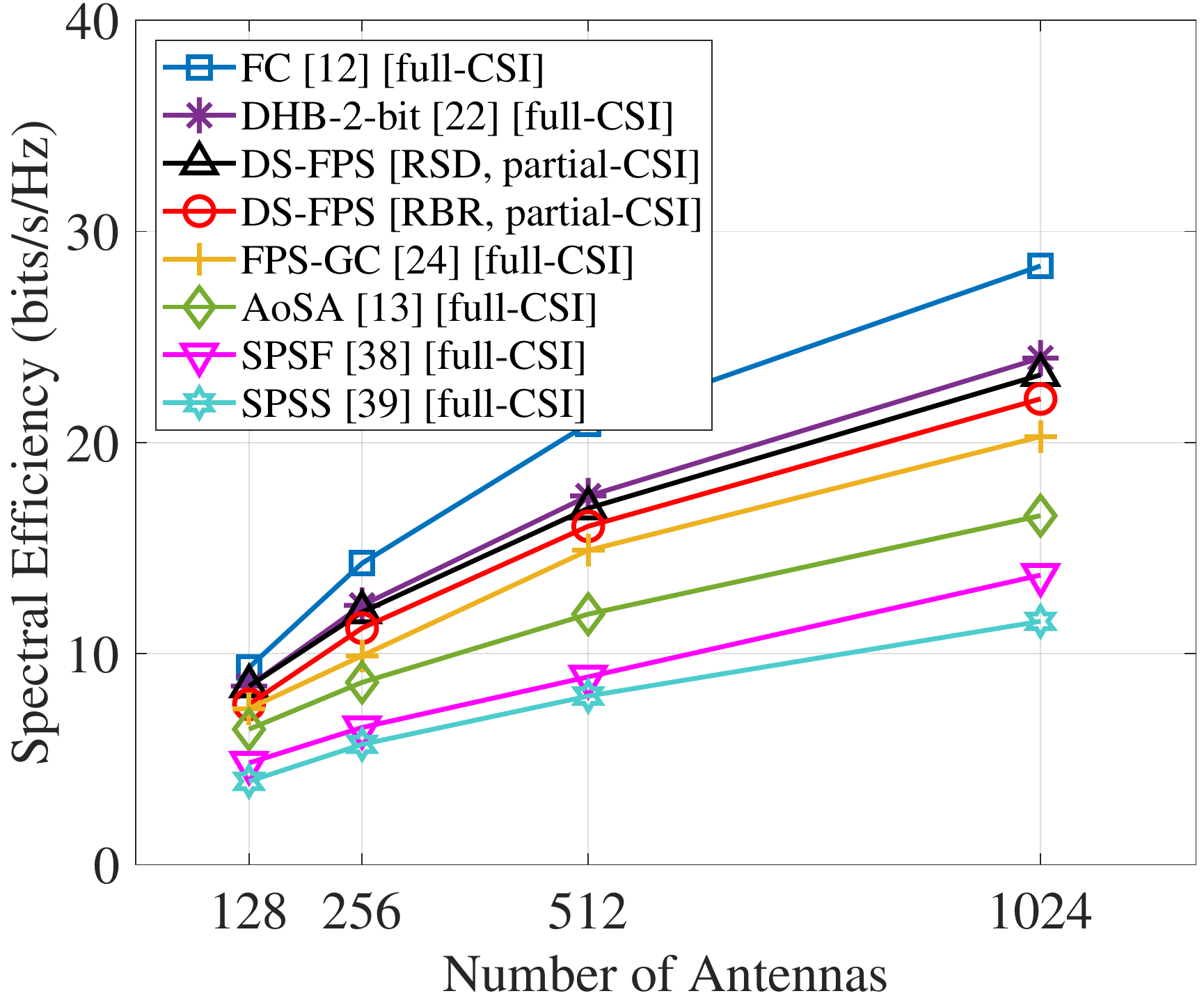}
			\captionsetup{font={footnotesize}}
			\caption{Spectral efficiency versus number of antennas. $D=40$m, $\rho=20$ dBm. Number of FPSs is $32$.}
			\label{fig_SE_Nt}
		\end{minipage}
		\begin{minipage}[t]{0.32\linewidth}
			\includegraphics[width = 1\linewidth]{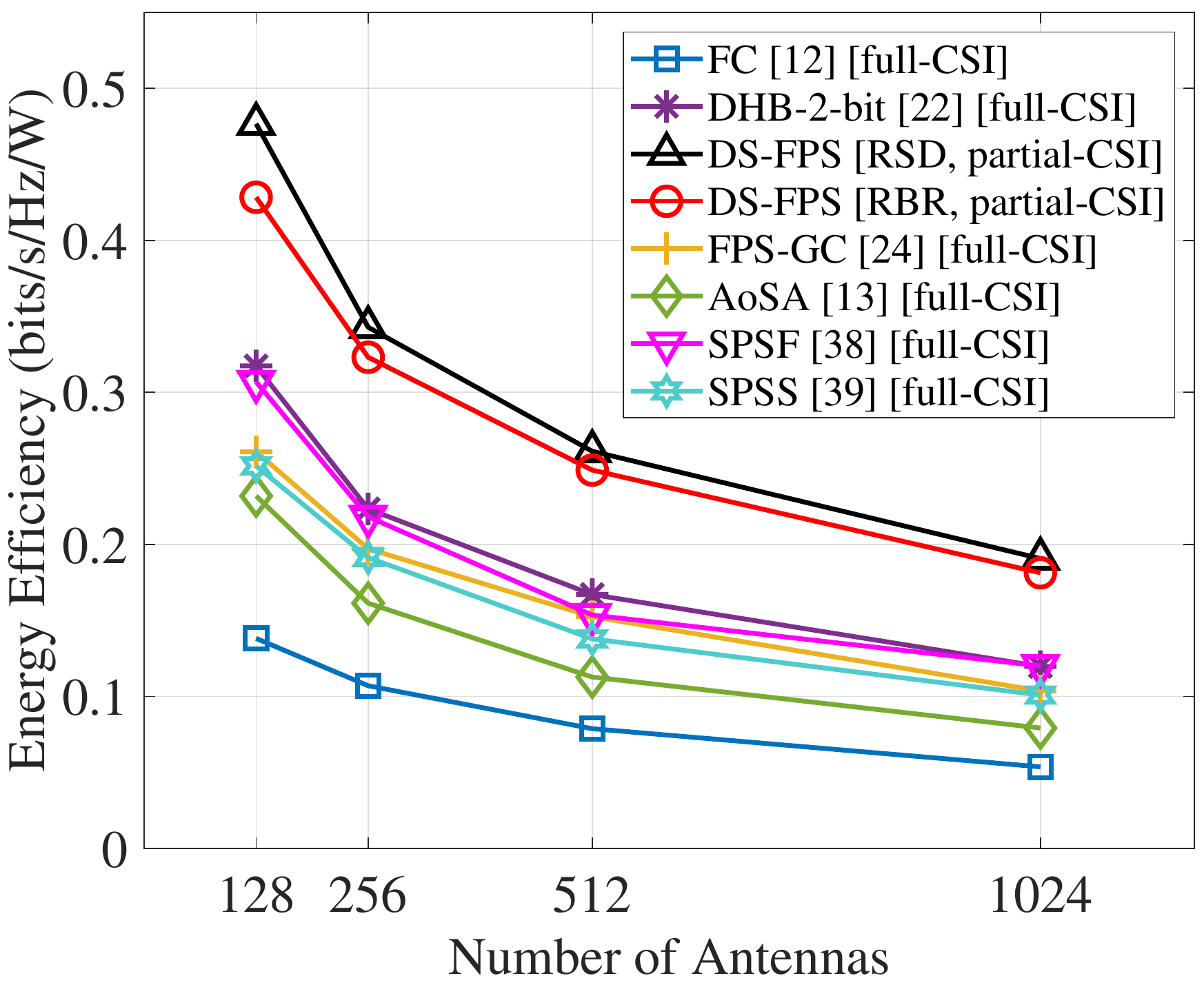}
			\captionsetup{font={footnotesize}}
			\caption{Energy efficiency versus number of antennas. $D=40$m, $\rho=20$ dBm. Number of FPSs is $32$.}
			\label{fig_EE_Nt}
		\end{minipage}
		\begin{minipage}[t]{0.34\linewidth}
			\includegraphics[width = 1\linewidth]{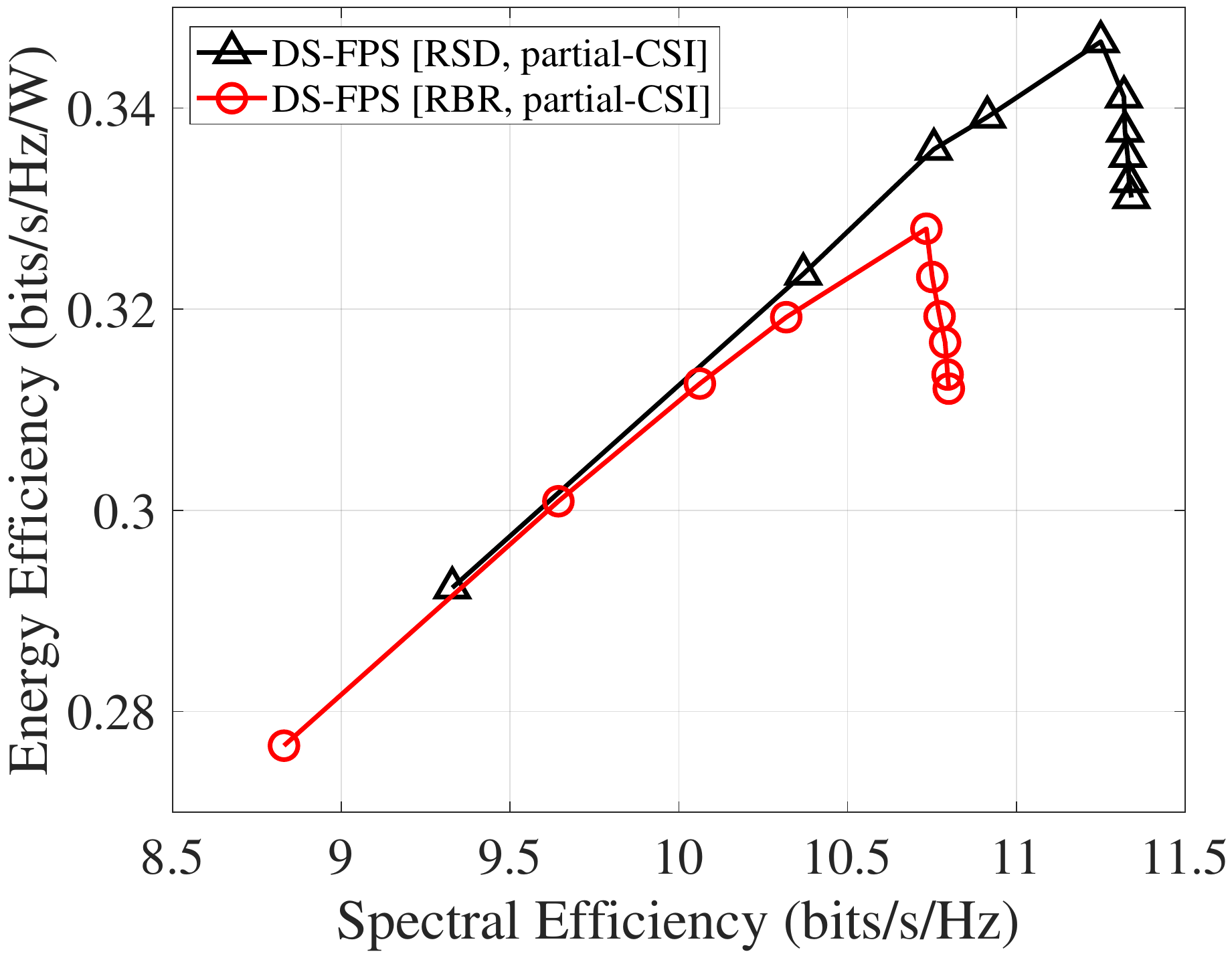}
			\captionsetup{font={footnotesize}}
			\caption{Energy efficiency versus spectral efficiency with different number of FPSs. $D=40$m, $\rho=20$ dBm, $N_t=N_r=256$.}
			\label{fig_SE_EE_FPS}
		\end{minipage}	
	\end{tabular}
	\vspace{-9.5mm}
\end{figure}

In Fig.~\ref{fig_SE_Nt} and Fig.~\ref{fig_EE_Nt}, we evaluate the spectral efficiency and energy efficiency versus number of antennas, respectively. On one hand, for all architectures, spectral efficiencies grow with the number of antennas, due to the higher array gain offered by more antennas. Moreover, the spectral efficiencies of the DS-FPS architecture with the RSD and RBR algorithms are always similar to the DHB with 2-bit phase shifter, which are higher than the FPS-GC, AoSA, SPSF, and SRPS architectures, as shown in Fig.~\ref{fig_SE_Nt}. On the other hand, with more antennas, the power consumption is higher such that the energy efficiencies of all architectures decrease, as shown in Fig.~\ref{fig_EE_Nt}. Particularly, for various numbers of antennas, the superiority on energy efficiency of the DS-FPS architecture stands out clearly.

In Fig.~\ref{fig_SE_EE_FPS}, we evaluate the impact of the number of FPSs on the spectral efficiency and energy efficiency of the proposed DS-FPS architecture. The number of FPSs from left to right is $12$, $16$, $20$, $24$, $32$, $40$, $56$, $72$, $96$, and $120$, respectively. The spectral efficiencies of the DS-FPS architecture with RSD and RBR algorithms increase with the number of FPSs, since the number of provided phases of the FPS network increases. When the number of FPSs exceeds $32$, the enhancement of spectral efficiency is negligible. By contrast, the energy efficiency of the DS-FPS architecture first increases and then decreases when the number of FPSs increases. The highest energy efficiency of the DS-FPS architecture is achieved with $32$ FPSs. The reason is that, when the number of FPSs is small, e.g., less than $32$, the increase of the number of FPSs provides higher spectral efficiency, which leads to higher energy efficiency. Differently, when the number of FPSs is too large, the increase of the spectral efficiency is negligible while the caused higher power consumption dominates the reduction of the energy efficiency. Therefore, $32$ FPSs are proper for the considered DS-FPS architecture to achieve both high spectral efficiency and energy efficiency.

\subsection{Impact of Partial and Inaccurate CSI}
\begin{figure}
	\centering
	\begin{tabular}{cc}
		\begin{minipage}[t]{0.32\linewidth}
			\includegraphics[width = 1\linewidth]{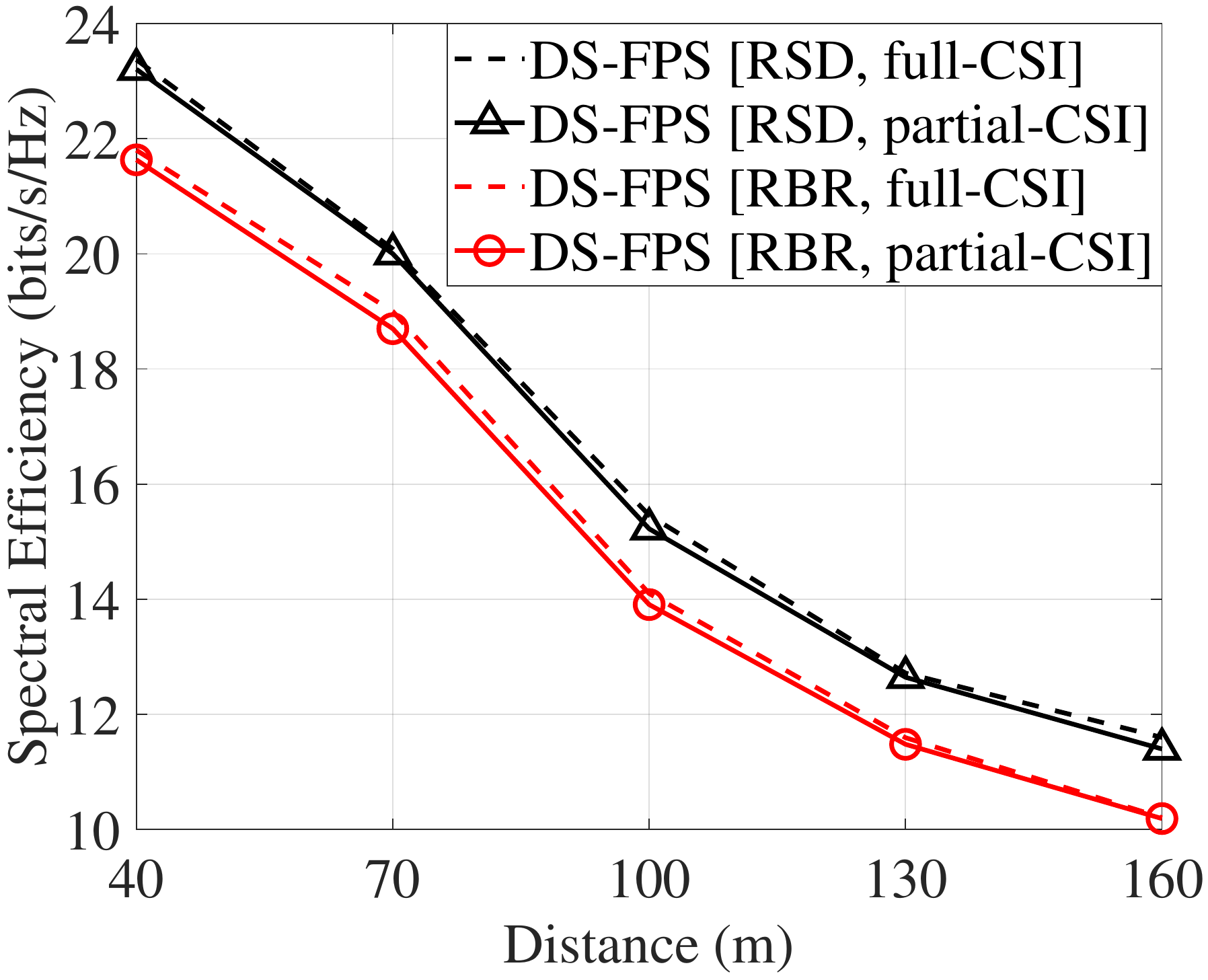}
			\captionsetup{font={footnotesize}}
			\caption{Spectral efficiency with full CSI and partial CSI. $N_t=N_r=1024$,~$Q=8$,~$\rho=20$ dBm.}
			\label{fig_SE_partial_CSI}
		\end{minipage}
		\begin{minipage}[t]{0.32\linewidth}
			\includegraphics[width = 1\linewidth]{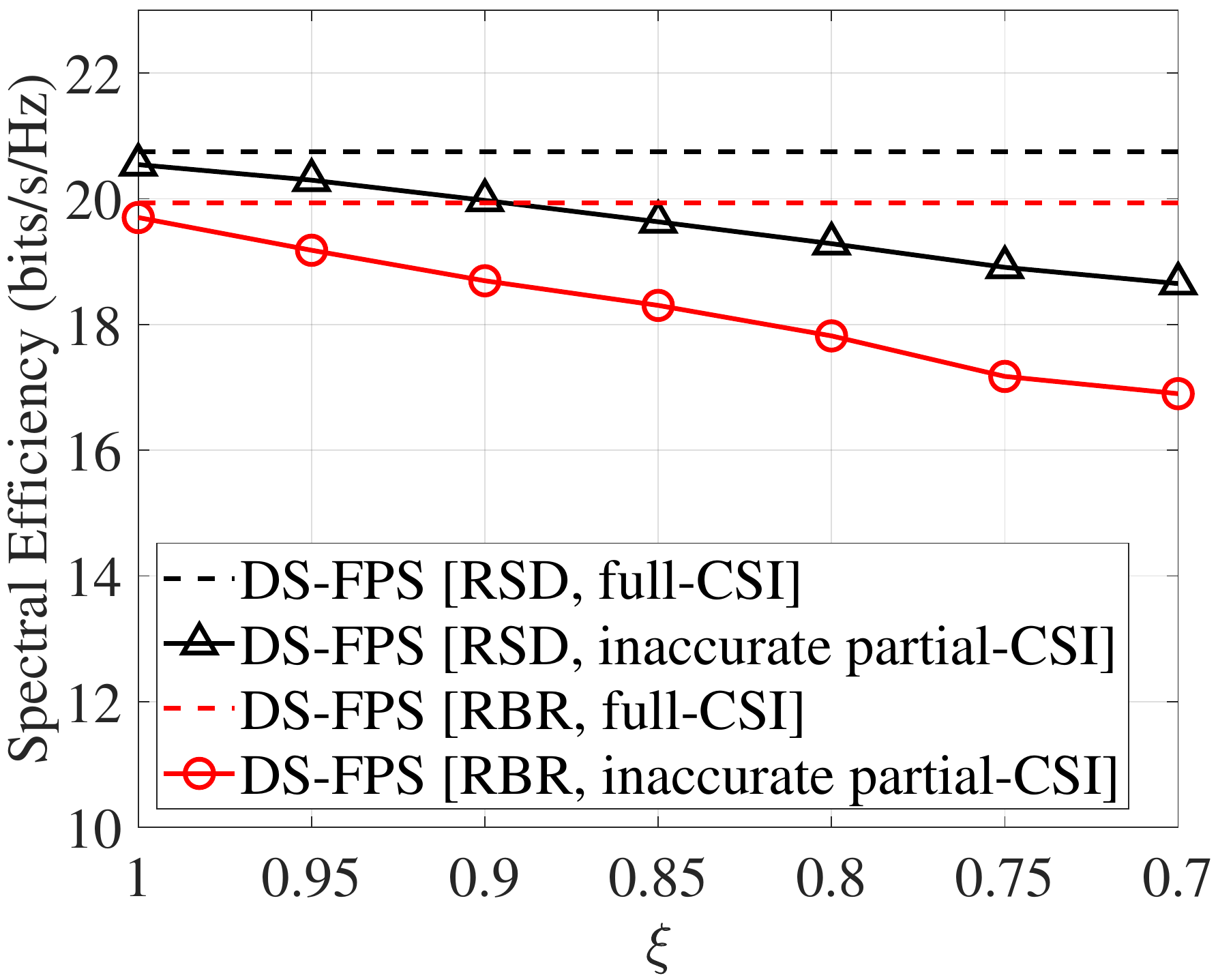}
			\captionsetup{font={footnotesize}}
			\caption{Spectral efficiency versus $\xi$. $N_t=N_r=1024$, $Q=8$, $\rho=20$ dBm, communication distance is $70$m.}
			\label{fig_SE_inaccurate_CSI}
		\end{minipage}
		\begin{minipage}[t]{0.315\linewidth}
			\includegraphics[width = 1\linewidth]{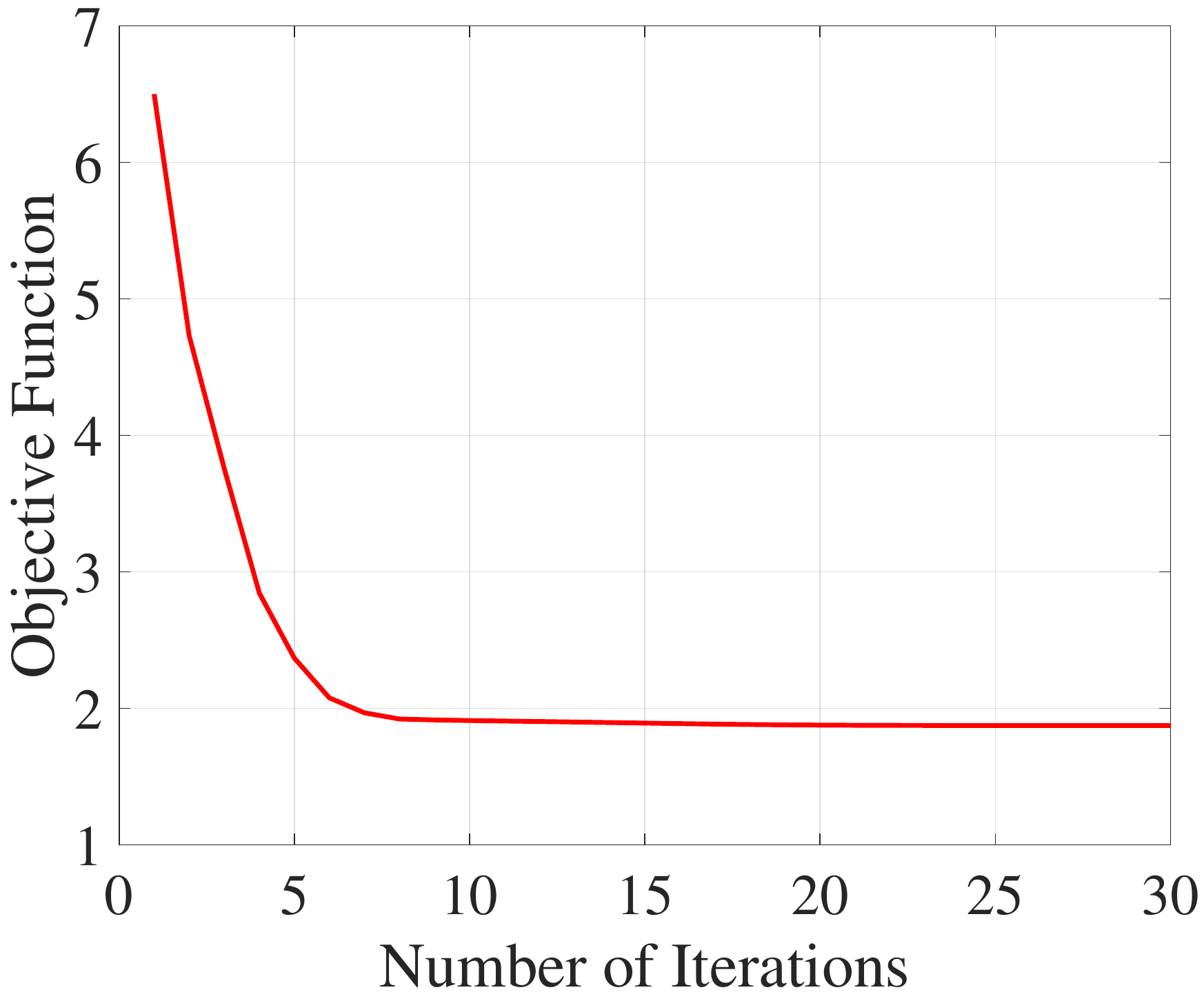}
			\captionsetup{font={footnotesize}}
			\caption{Convergence of the RBR algorithm. $N_t=N_r=1024$, $Q=8$. Communication distance is $40$m.}
			\label{fig_convergence_RBR}
		\end{minipage}	
	\end{tabular}
	\vspace{-9.5mm}
\end{figure}

We evaluate the impact of partial and inaccurate CSI on the spectral efficiency of the DS-FPS architecture. The proposed RSD and RBR algorithms are designed for the case of partial CSI. With the following adjustment, the RSD and RBR algorithms can also work for the case of full CSI. $\bm{\bar\Lambda}[k]\textbf{A}_{t}[k]^H$ in~\eqref{design_problem_1} should be replaced by the full channel matrix $\textbf{H}[k]$ and the associated steps of the RSD and RBR algorithms need to be changed accordingly. 
Fig.~\ref{fig_SE_partial_CSI} presents the spectral efficiency versus communication distance of DS-FPS architecture with partial CSI and full CSI. With partial CSI, both the proposed RSD and RBR algorithms can achieve the similar spectral efficiency of the case with full CSI, e.g., 97\% with $40$m distance.

In Fig.~\ref{fig_SE_inaccurate_CSI}, we evaluate the spectral efficiency of the RSD and RBR algorithms under inaccurate partial CSI at TX and RX, i.e., the known $\bm{\bar{\Lambda}}[k]$, $\textbf{A}_{t}[k]$, and $\textbf{A}_{r}[k]$ are inaccurate. We use $\xi\in[0,1]$ to represent the level of accuracy. The inaccurate $\bm{\bar{\Lambda}}[k]$ can be represented as
\begin{align}
\bm{\bar{\Lambda}}_{\hbar}[k]=\xi\bm{\bar{\Lambda}}[k]+\sqrt{1-\xi^2}\textbf{E}_1\odot\bm{\bar{\Lambda}}[k],
\label{eq_inaccurate_CSI}
\end{align}
where $\textbf{E}_1$ denotes the error matrix with elements following the distribution of i.i.d $\mathcal{N}(0,1)$. For the inaccurate $\textbf{A}_{t}[k]$ and $\textbf{A}_{r}[k]$, the expressions are similar to~\eqref{eq_inaccurate_CSI} while the inaccuracy is enforced on the phase since the amplitude of the elements of these two matrices should be fixed as one, as stated in~\eqref{steering_vector_UPA}. Fig.~\ref{fig_SE_inaccurate_CSI} shows the spectral efficiency of the RSD and RBR algorithms by varying the value of $\xi$. The dash lines represent the case of full CSI which is accurate and not related to $\xi$. Compared to the case of full CSI, the spectral efficiency loss of the RSD and RBR algorithms under inaccurate partial CSI grows with the smaller $\xi$. Note that the spectral efficiency loss of both RSD and RBR algorithms maintains less than 15\% when $\xi\geq 0.7$, which is acceptable and reveals that the proposed RSD and RBR algorithms are robust to CSI error.

\subsection{Convergence and Computational Complexity Analysis}

Fig.~\ref{fig_convergence_RBR} shows the convergence performance of the RBR algorithm with partial CSI. We evaluate the objective function of the RBR algorithm, i.e., $\frac{1}{K}\sum\nolimits_{k=1}^{K}\lVert \textbf{P}[k]-\textbf{S}\textbf{F}\textbf{D}[k]\rVert_{F}^2$, versus the number of iterations. We run the RBR algorithm 1000 times to take the average. The simulation results show that the RBR algorithm converges fast with about 15 iterations.

As listed in Table~\ref{table_computational_complexity}, we analyze the computational complexities of the RSD and RBR algorithms. In THz UM-MIMO systems, the number of antennas $N_t$ is usually very large, e.g., 1024. The number of RF chains $L_t$, the number of data streams $N_s$, the number of multipath of channel $N_p$, and the number of FPSs $Q$ within each RF chain are small, e.g., $L_t=N_s=4$, $N_p=6$, and $Q=8$ in our simulations. Without loss of generality, we have ${\mathcal{O}}(L_t)\approx{\mathcal{O}}(N_s)\approx{\mathcal{O}}(N_p)\approx{\mathcal{O}}(Q)\ll {\mathcal{O}}(N_t)$. The RSD algorithm does not involve iterations, whose overall computational complexity is $ {\mathcal{O}}(KN_tN_p^3L_tQ)$. The RBR algorithm includes iterations and we denote $M$ as the number of iterations. The overall computational complexity of RBR algorithm is ${\mathcal{O}}(KMN_tL_tQ)$. Specifically, for the calculation of $\textbf{P}[k]$, RBR algorithm needs to calculate the $N_s$-truncated SVD of $\bm{\bar\Lambda}[k]\textbf{A}_t[k]^H$ to obtain the first $N_s$ columns of the right singular matrix, whose complexity is ${\mathcal{O}}((N_t+N_p)N_s^2)$~\cite{8438554}. Through the simulation in Fig.~\ref{fig_convergence_RBR}, we observe that, the RBR algorithm can converge with about $15$ iterations. Therefore, $M$ is usually much smaller than $N_p^3$. Consequently, although the RBR algorithm has lower spectral efficiency than the RSD algorithm as analyzed before, its computational complexity is also lower than the RSD algorithm.

\begin{table}
	\centering
	\footnotesize
	\captionsetup{font={footnotesize}}
	\caption{Computational complexity analysis of the proposed RSD and RBR algorithms.}
	\begin{tabular}{ll} 
		\hline
		\multicolumn{2}{c}{RSD algorithm}\\
		\hline Operation&Complexity\\
		Design $\textbf{S}_i$ via maximizing~\eqref{expression_u} for $i=1$, $2$, ..., $N_t$ &${\mathcal{O}(KN_tN_p^3L_tQ)}$\\
		Calculate $\textbf{D}[k]$ through~\eqref{solution_D_RSD}, for $k=1$, $2$, ..., $K$&${\mathcal{O}}(K(N_tL_t^2Q+(N_p+L_t)N_s^2))$\\
		Normalize each $\textbf{D}[k]$ as $\textbf{D}[k]=\frac{\rho}{\sum_{k=1}^{K}\lVert\textbf{S}\textbf{F}\textbf{D}[k]\rVert_F^2}\textbf{D}[k]$&$\mathcal{O}(KN_tL_tN_s)$\\		
		\textbf{Overall}  &$\bm{ {\mathcal{O}}(KN_tN_p^3L_tQ)}$\\
		\hline \multicolumn{2}{c}{RBR algorithm}\\
		\hline Operation&Complexity\\
		Calculate $\bm{\bar\Lambda}[k]\textbf{A}_t[k]^H$, for $k=1$, $2$, ..., $K$&$\mathcal{O}(KN_tN_p^2)$\\
		Calculate each $\textbf{P}[k]$ through $N_s$-truncated SVD of $\bm{\bar\Lambda}[k]\textbf{A}_t[k]^H$, for $k=1$, $2$, ..., $K$ &$\mathcal{O}(K(N_t+N_p)N_s^2)$\\
		Calculate each $\textbf{D}[k]$ via~\eqref{solution_D_RBR}&$\mathcal{O}(K(N_tL_tN_s+(L_t+N_s)N_s^2))$\\
		Design $\textbf{S}_i$ via~\eqref{solution_S_RBR} for $i=1$, $2$, ..., $N_t$&$\mathcal{O}(KN_tL_tQ)$\\
		Normalize each $\textbf{D}[k]$ as $\textbf{D}[k]=\frac{\rho}{\sum_{k=1}^{K}\lVert\textbf{S}\textbf{F}\textbf{D}[k]\rVert_F^2}\textbf{D}[k]$&$\mathcal{O}(KN_tL_tN_s)$\\
		\textbf{Overall (M iterations)}	&$\bm {{\mathcal{O}}(KMN_tL_tQ)}$\\
		\hline
	\end{tabular}
	\label{table_computational_complexity}
	\vspace{-9.5mm}
\end{table}

The computational complexities of the algorithm in~\cite{7913599} for FC, the algorithm in~\cite{9110865} for DS, the SIC algorithm~\cite{7445130} for AoSA, and the FPS-AltMin algorithm~\cite{8310586} for FPS-GC are $\mathcal{O}(KN_t^3)$, $\mathcal{O}(KN_t^3L_t)$, $\mathcal{O}(KN_t^2L_t^2)$, and $\mathcal{O}(KN_tL_t^2N_s^2Q\cdot{\rm log}_2(KN_tL_t^2Q))$, respectively.
The computational complexity of the RBR algorithm is much lower than these existing algorithms. Although the RSD algorithm has higher complexity than the RBR, it has higher spectral efficiency as well.
The digital beamforming of the algorithms for SRPS and SPSF architectures is the same with computational complexity $\mathcal{O}(N_t^2)$~\cite{8295113,8382230}. The analog beamforming of these two algorithms are based on the first $L_t$ columns of the right singular matrix of channel while with different strategy about the turning off of the antennas. The computational complexity of the analog beamforming of these two algorithms is proportional to $\mathcal{O}(N_t)$. Hence, the computational complexity of the algorithms for SPSF and SRPS architectures is proportional to $\mathcal{O}(N_t^2)$, which is higher than the RSD and RBR algorithms in this work, i.e., linearly related with $N_t$.

\section{Conclusion}
\label{section_conclusion}
In this paper, we have proposed an energy-efficient DS-FPS architecture for THz hybrid beamforming systems, using low-cost FPSs. Moreover, to account for practical partial CSI, we have proposed an RSD algorithm to design the hybrid beamforming matrices for the DS-FPS architecture. To further reduce the computational complexity of the RSD algorithm, an RBR algorithm has been developed.
Extensive simulation results have shown that the DS-FPS architecture achieves remarkably higher energy efficiency than the existing architectures. Moreover, by using the RSD and RBR algorithms, the DS-FPS architecture based on partial CSI can achieve $97\%$ spectral efficiency of that with full CSI. When the partial CSI is inaccurate with errors, the spectral efficiency loss compared to full CSI is still less than $15\%$, which is acceptable and reveals that the proposed RSD and RBR algorithms are robust to the incompleteness and inaccuracy of CSI. Furthermore, we have analytically found that the computational complexity of the RBR algorithm is linearly related with the number of antennas, which is much lower than the existing algorithms. Compared to the RBR algorithm, the RSD algorithm yields higher spectral efficiency, at the cost of increased computational complexity.

\bibliographystyle{IEEEtran}
\bibliography{IEEEabrv,reference}
\end{document}